\makeatletter

\newcommand{\Rmnum}[1]{\expandafter\@slowromancap\romannumeral #1@}
\makeatother

\documentclass{emulateapj}
\usepackage{latexsym,bm}
\usepackage{textcmds}
\usepackage{amsmath}
\usepackage{rotating}

\shorttitle{Lensed star-forming galaxy}
\shortauthors{Bian et al.}

\begin{document}



\title{LBT/LUCIFER Observations of the $z\sim2$ Lensed Galaxy J0900+2234\altaffilmark{1}}


\author{Fuyan Bian\altaffilmark{2}, Xiaohui Fan\altaffilmark{2}, Jill Bechtold\altaffilmark{2}, 
Ian D. McGreer\altaffilmark{2}, Dennis W. Just\altaffilmark{2}, David J. Sand\altaffilmark{3,9},
Richard F. Green\altaffilmark{4}, David Thompson\altaffilmark{4}, 
Chien Y. Peng\altaffilmark{5}, Walter Seifert\altaffilmark{6}, Nancy Ageorges\altaffilmark{7}, 
Marcus Juette\altaffilmark{8},
Volker Knierim\altaffilmark{8}, Peter Buschkamp\altaffilmark{7}}

\altaffiltext{1}{Based on data acquired using the Large Binocular Telescope
(LBT). The LBT is an international collaboration among institutions
in the United States, Italy and Germany. LBT Corporation
partners are: The University of Arizona on behalf of the Arizona
university system; Istituto Nazionale di Astrofisica, Italy; LBT
Beteiligungsgesellschaft, Germany, representing the Max-Planck
Society, the Astrophysical Institute Potsdam, and Heidelberg
University; The Ohio State University, and The Research Corporation,
on behalf of The University of Notre Dame, University of
Minnesota and University of Virginia.}
\altaffiltext{2}{Steward Observatory, University of Arizona, 933 N Cherry Ave., Tucson, AZ 85721, USA}
\altaffiltext{3}{Harvard-Smithsonian Center for Astrophysics, 60 Garden Street, Cambridge, MA 02138, USA} 
\altaffiltext{4}{Large Binocular Telescope Observatory, University of Arizona, 933 N. Cherry Ave., Tucson, AZ 85721, USA}
\altaffiltext{5}{Herzberg Institute of Astrophysics, National Research Council of Canada
5071 West Saanich Road, Victoria, British Columbia, V9E 2E7, Canada}
\altaffiltext{6}{Landessternwarte K\"{o}nigstuhl, Zentrum f\"{u}r Astronomie Heidelberg, K\"{o}nigstuhl 12, 69117 Heidelberg,
Germany}
\altaffiltext{7}{Max Planck Institut f\"ur extraterrestrische Physik, Giessenbachstrasse 1, 85748 Garching, Germany}
\altaffiltext{8}{Astronomisches Institut, Ruhr-Universit\"at, Universit\"atstrasse 150, D-44780 Bochum, Germany}
\altaffiltext{9}{Harvard Center for Astrophysics and Las Cumbres Observatory Global Telescope Network Fellow.}



\begin{abstract}
We present rest-frame optical images and spectra of the 
gravitationally lensed, star-forming galaxy J0900+2234 ($z=2.03$).
The observations were performed with the newly commissioned LUCIFER1 
near-infrared (NIR) instrument mounted on the Large Binocular Telescope (LBT).
We fitted lens models to the rest-frame optical images and found the galaxy has
an intrinsic effective radius of $7.4\pm0.8$~kpc with a lens magnification factor
of about 5 for the A and B components.
We also discovered a new arc belonging to another lensed high-z source 
galaxy, which makes this lens system
a potential double Einstein ring system. Using the high S/N rest-frame optical
spectra covering $H$+$K$ band, we detected H$\beta$, [\ion{O}{3}], H$\alpha$,
[\ion{N}{2}], and [\ion{S}{2}] emission lines. Detailed physical properties
of this high-z galaxy were derived. The extinction towards
the ionized \ion{H}{2} regions ($E_g(B-V)$) was computed from the flux ratio
of H$\alpha$ and H$\beta$ and appears to be much higher than that towards
stellar continuum ($E_s(B-V)$), derived from the optical and NIR broad band 
photometry fitting. The metallicity was estimated using N2 and O3N2 indices.
It is in the range of $\frac{1}{5}-\frac{1}{3}$ solar abundance, which 
is much lower than the typical $z\sim2$ star-forming galaxies.
From the flux ratio of [\ion{S}{2}]$\lambda6717$ and [\ion{S}{2}]$\lambda6732$,
we found that the electron number density of 
the \ion{H}{2} regions in the high-z galaxy were $\simeq1000$~cm$^{-3}$, consistent with other $z\sim2$ galaxies but 
much higher than that in local \ion{H}{2} regions. 
The star-formation rate was estimated via the $H\alpha$ luminosity, 
after correction for the lens magnification, to be about $365\pm69~M_{\sun}~yr^{-1}$.
 Combining the FWHM of H$\alpha$ emission 
lines and the half-light radius, we found the dynamical mass of the lensed galaxy is 
$(5.8\pm0.9)\times10^{10}~M_{\sun}$. The gas mass is $(5.1\pm1.1)\times10^{10}~M_{\sun}$
from the H$\alpha$ flux surface density by using global Kennicutt-Schmidt Law, indicating 
a very high gas fraction of $0.79\pm0.19$ in J0900+2234.
\end{abstract}





\section{Introduction}
The redshifts between $1<z<3$ mark the time of peak cosmic star formation
\citep{madau98} and quasar activity \citep{fan01}.
The current-day Hubble sequence was being built up, and a large 
fraction of present-day stars were formed \citep{dickinson03}.
Observations of galaxies in this redshift 
range provide a direct picture of the formation and evolution of 
galaxies and the assembly of their central supermassive black holes. 

By obtaining rest-frame optical spectra, we can perform detailed studies of the star formation history, 
as well as properties of the ISM and stellar populations 
 in galaxies. However, well-known diagnostic optical 
emission lines are shifted into the near infrared (NIR) for $z=1-3$. Since high-z 
galaxies (e.g., Lyman Break Galaxies or LBGs) 
are faint, it is difficult to obtain high signal to noise ratio (S/N) NIR spectra with 
current facilities, 
especially for weak emission and absorption lines. 
Therefore, many previous NIR spectroscopic studies have been focused on the strong
[\ion{O}{3}] or H$\alpha$ 
 emission lines \citep[e.g.,][]{pett01, erb03, erb06b, erb06c}, from which only limited
inferences can be drawn. 

One method of addressing this problem is to observe high-z star-forming galaxies which have been 
gravitationally lensed by foreground 
massive galaxies or clusters. Lensing can boost the observed flux of high redshift galaxies 
by several tens or more, so that high S/N spectroscopy in the NIR becomes feasible. 
One of the first lensed galaxies to be studied in this way was the serendipitously 
discovered MS1513-cB58 at $z=2.73$ \citep{yee96, bechtold97, teplitz00}.
Subsequently, several groups began systematic searches for strongly lensed high redshift 
galaxies towards clusters of galaxies \citep[]{sand05, richard08} and red galaxies in the 
Sloan Digital Sky Survey \citep[SDSS,][]{york00} 
images \citep[e.g.,][]{belo09,kubo09}. These searches are finding lensed star-forming 
galaxies at redshifts $z>2$ \citep[e.g.,][]{alla07,lin09, dieh09,koes10} which can then 
be followed-up in the NIR, or rest-frame optical and UV 
\citep[e.g.,][]{smai07, hain09, fink09,quid09,quid10,pett10}.

The object of this paper is the 
lensed system J0900+2234 \citep[z=2.03;][]{dieh09} which was discovered in SDSS images,
and then confirmed with follow-up deep optical $g$, $r$, $i$ imaging and spectroscopy.
The lens galaxy is at $z = 0.49$ based on SDSS spectroscopy. 
In this paper, we report $J$, $H$, $Ks$ imaging and $H$+$K$ spectroscopy 
of J0900+2234 with the newly commissioned LUCIFER instrument \citep{mand08,Ager10}
at the Large Binocular Telescope \citep[LBT;][]{hill08}.
We detected the nebular emission lines H$\beta$, [\ion{O}{3}]$\lambda\lambda$4959,5007, 
H$\alpha$, [\ion{N}{2}]$\lambda$6583, and [\ion{S}{2}]$\lambda\lambda$6717,6732 
with simultaneous $H$+$K$ band spectral coverage from 1.40-2.20$\mu m$. 
The wide simultaneous wavelength range coverage not only improves observing efficiency, 
but also helps to reduce the measurement uncertainties of the line ratios (e.g., H$\alpha$/H$\beta$)
resulting from the flux calibration and slit losses variation between different exposures.

This paper is organized as follows: 
In \S~\ref{observation}, we describe the observations and reduction of the LUCIFER data.
In \S~\ref{measurements}, we describe the photometry, reconstruction of the source with 
lens modeling and line flux measurements. 
In \S~\ref{lensmodel}, we construct a simple lensing model using the J0900+2234 NIR
imaging. 
In \S~\ref{physics}, we report the detailed physical properties of
J0900+2234 and discuss the systematic uncertainties of the 
physical property measurements. In \S~\ref{conclusion}, our main results are summarized.
Throughout this paper, we use a cosmology with $H_0=70$~km~s$^{-1}$~Mpc$^{-1}$, $\Omega_{\rm M}=0.3$,
and $\Omega_{\Lambda}=0.7$. The magnitudes are all AB magnitudes.

\section{Observations and Data Reduction}\label{observation}
\subsection{LUCIFER}

LUCIFER1 was built by a collaboration of five German institutes. It is the first of a pair of NIR
imagers/spectrographs for the LBT. It is mounted on the bent Gregorian focus of the left primary mirror. The
wavelength coverage is from 0.85 to 2.4 $\mu$m ($zJHK$ bands) in imaging, long-slit and multiobject
spectroscopy modes . The detector is a Rockwell HAWAII-2 Hd-CdTe $2048\times2048$ pixel$^2$ array
with a pixel size of 18.5 $\mu$m. The quantum efficiency (QE) is about 75\% in the $H$ and $K$ bands.
In the currently available seeing limited mode, the pixel scales are 0.25$^{\prime\prime}$ and 
0.12$^{\prime\prime}$ for N1.8 \& N3.75 
cameras, respectively. The commissioning finished in early December 2009 and has been immediately
followed by 12 nights of scientific demonstration observing. The instrument has been in regular science
operation since mid-December 2009.



\subsection{LUCIFER Imaging \& Spectroscopy}
J0900+2234 was observed with LUCIFER1 on the LBT
on 6 January 2010 in imaging mode and on 13 February 2010 in long slit 
spectroscopy mode (Table~\ref{log}). 
In the imaging mode,
the N3.75 camera was used; the 18.5 $\mu$m pixels correspond to 0.12$^{\prime\prime}$ on 
the sky, for a 4.0$^{\prime}\times$4.0$^{\prime}$ field-of-view. The seeing
was 0.7$^{\prime\prime}$ with thin clouds. 
For each filter ($J$, $H$, and $Ks$), we obtained ten randomly dithered exposures of 60s each. 
For the spectroscopic observations, the seeing was 0.5-0.6$^{\prime\prime}$
and the conditions were photometric. The spectra were taken with N1.8 camera, yielding a plate 
scale of 0.25$^{\prime\prime}$ per pixel. 
An order separation filter and the 200 l/mm $H$+$K$ grating were used to cover the wavelength range
from 1.40 to 2.20 $\mu$m simultaneously. 
A 1.0$^{\prime\prime}$ by 3.9$^{\prime}$ slit was used, resulting in spectral resolution of 16~{\AA}.
The total exposure time was $24\times 300$ seconds with the telescope nodded $\sim7^{\prime\prime}$
along the slit between each exposure. The instrument was rotated to PA=86.57$^{\circ}$ in order
to place the two brightest lensed components in the slit. Dark exposures and internal quartz 
lamp flats were obtained in the afternoon, and an Ar lamp was observed for wavelength 
calibration during the night. 

\subsection{Data Reduction}

\subsubsection{Imaging Reduction}
The imaging data were reduced using standard IDL routines.
The dark frames were median combined to create a master dark frame. A 
super-sky flat was constructed from combining the science frames after 
dark subtraction. The science images were dark subtracted, and then 
divided by the super-sky flat to correct for the detector response. 
The science images were registered to the SDSS-DR6 catalog using SCAMP 
\citep{bert06}
for astrometric calibration; the residuals are less than 0.1$^{\prime\prime}$ RMS. 
Finally, the sky background was subtracted from the science frames
which were then co-added using SWarp \citep{bert02}.
Photometric zeropoints were determined from 2MASS stars in the field-of-view.

\begin{figure}
\center
\includegraphics[scale=.31]{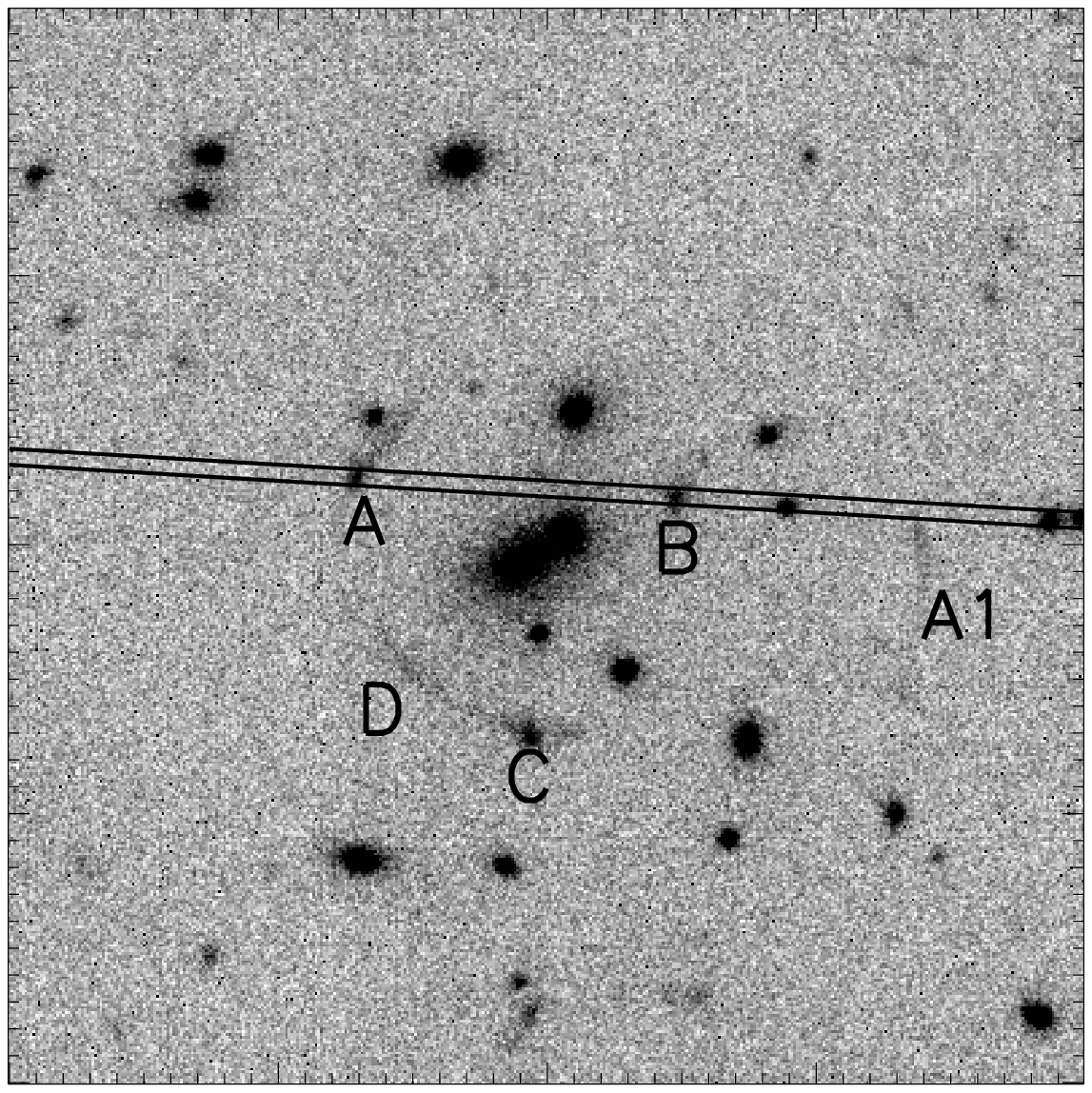}
\includegraphics[scale=.31]{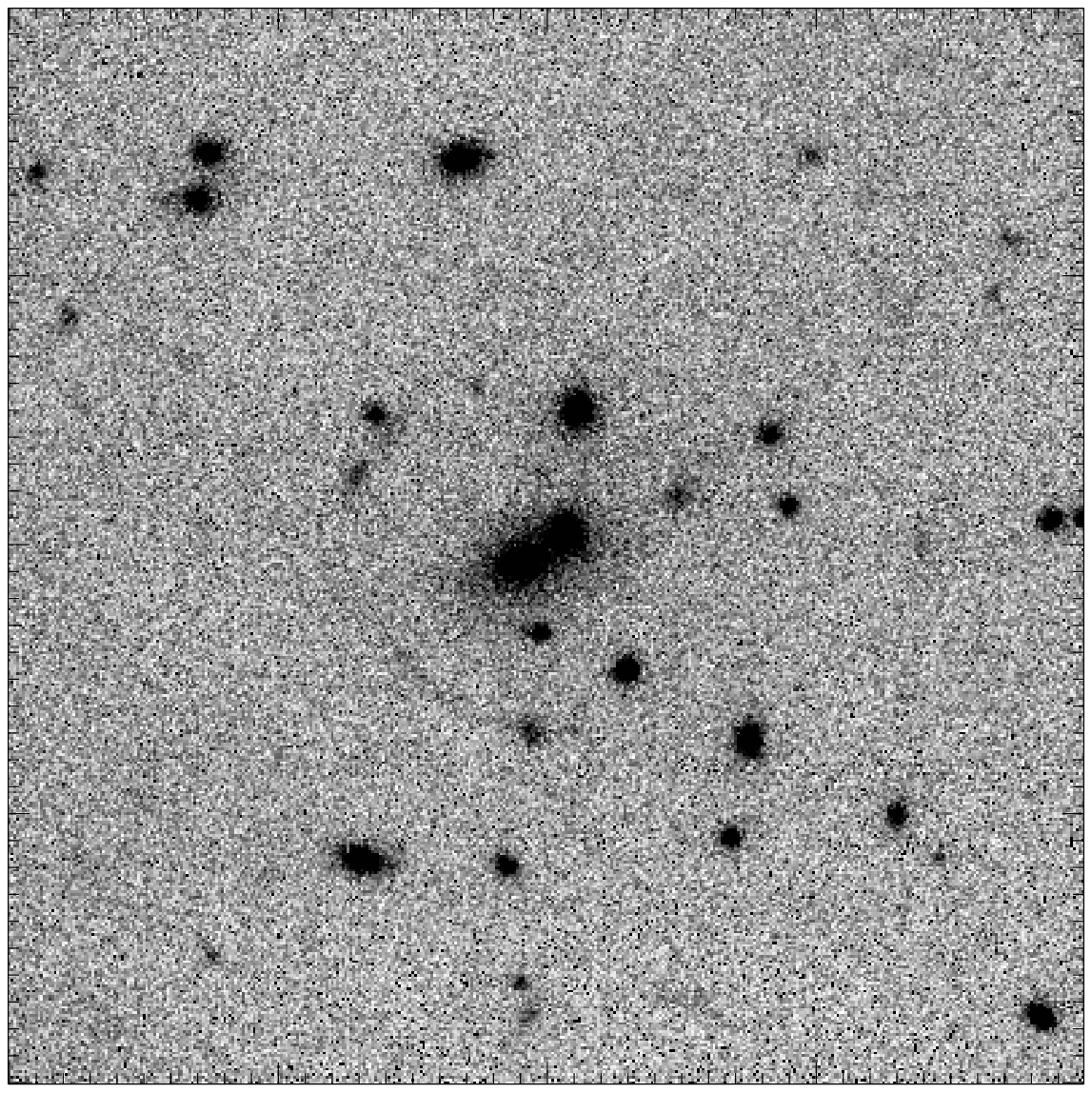}
\includegraphics[scale=.31]{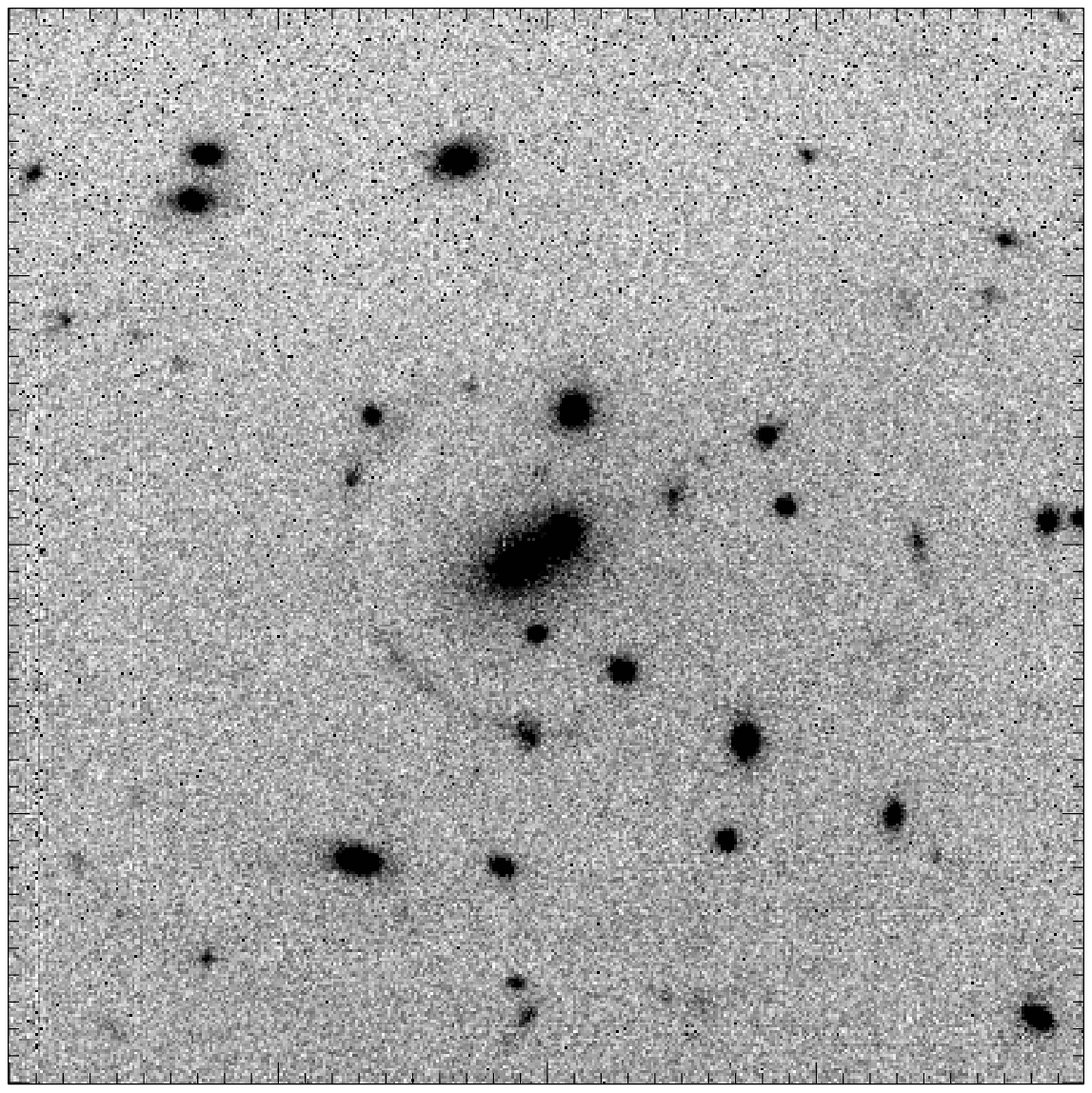}
\caption{$J$, $H$, and $Ks$ band images of J0900+2234 from LUCIFER, North is up, and
east is left. Each image shown has $40^{\prime\prime}\times40^{\prime\prime}$
The previously discovered A, B, C, and D components are labeled in the $J$-band image.
The A, B, and D components are images of the star-froming galaxy at $z\sim2$.
Our follow-up spectroscopy shows that C is not an image of the same galaxy. 
The newly discovered A1 arc from another lensed galaxy
is also labeled in the image. \label{imgs}} 
\end{figure}

The fully reduced $J$, $H$, $Ks$-band images are shown in Fig.~\ref{imgs}. In addition to the lensed galaxy
(components A, B, C, and D in Fig.~\ref{imgs}) found by \citet{dieh09}, we discovered another 
component, A1. The color of A1 suggests that it is also a high-redshift 
galaxy image lensed by the foreground red galaxy, but not the same galaxy 
appearing in images as A, B, C or D (see details in section~\ref{phot}).

\subsubsection{Spectroscopy Reduction}
\begin{figure}
\center
\plottwo{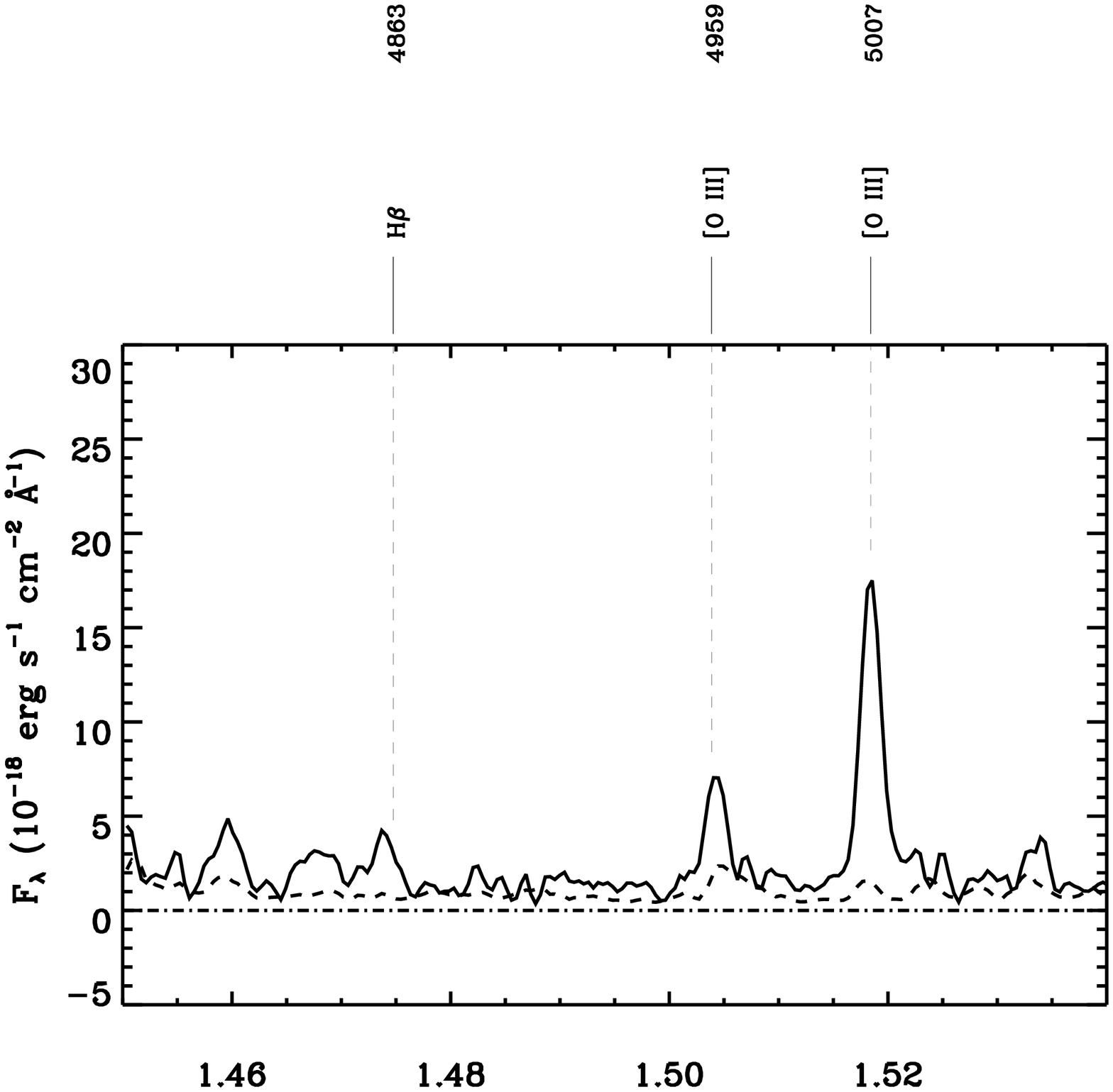}{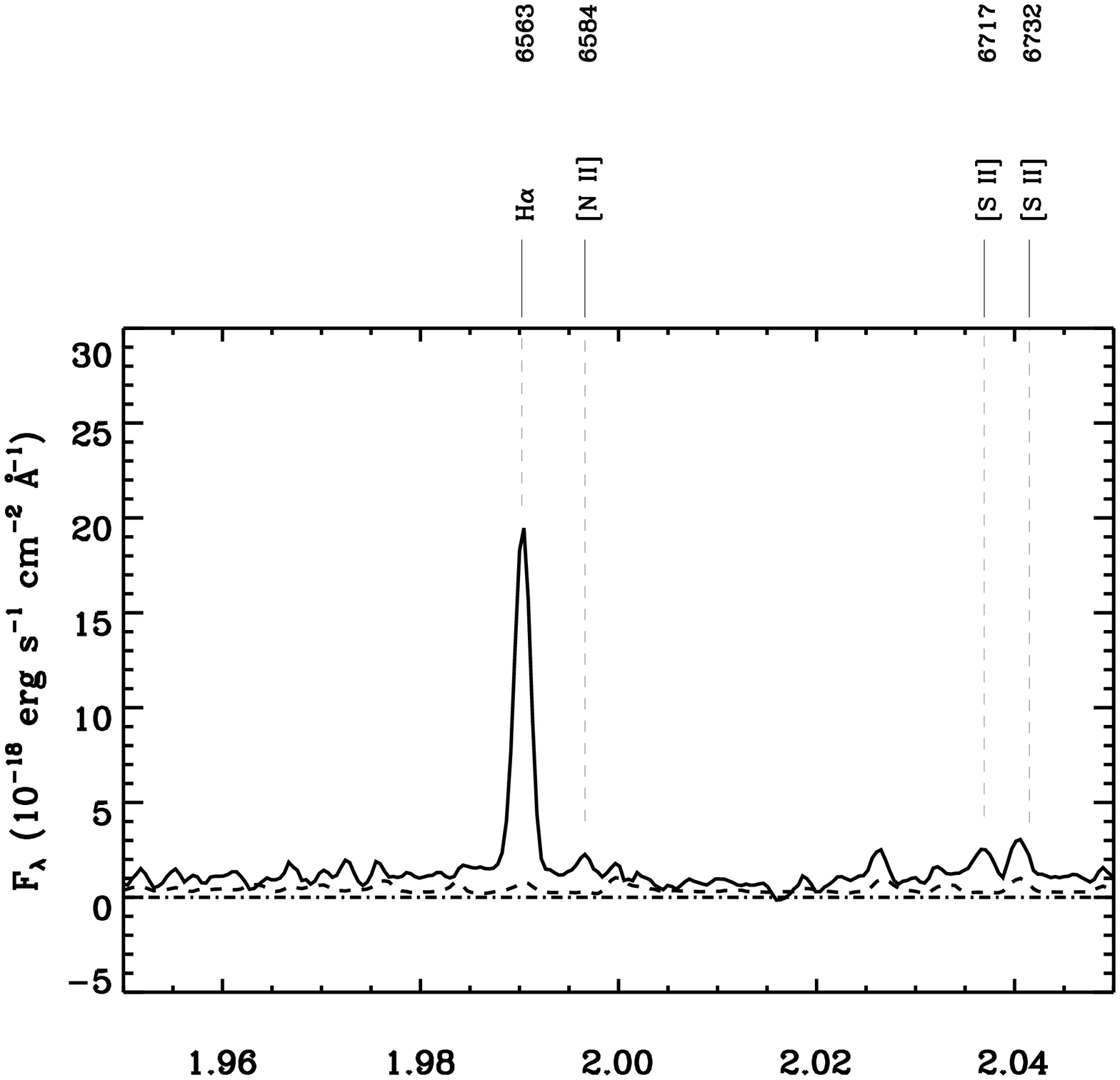}
\plottwo{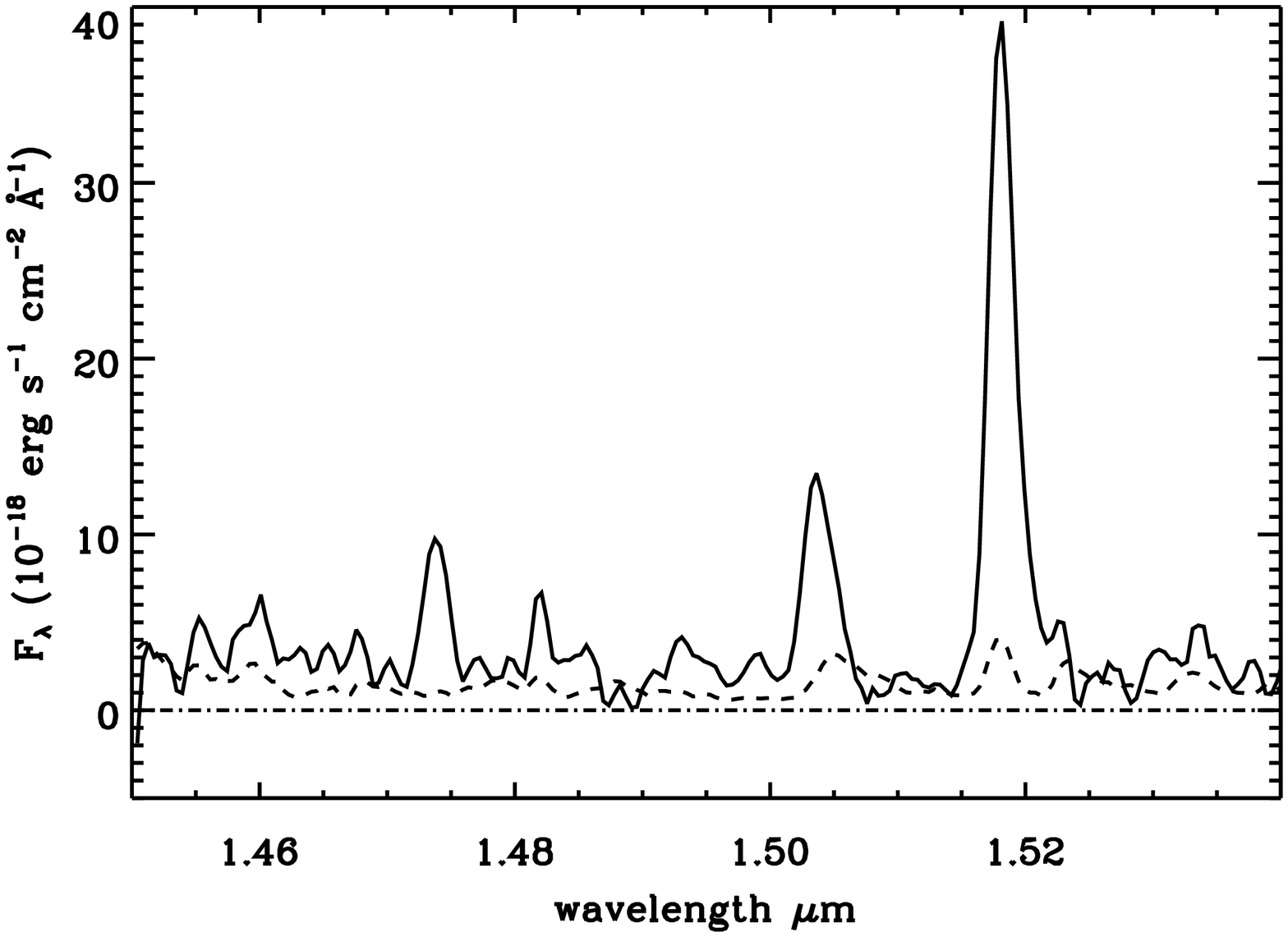}{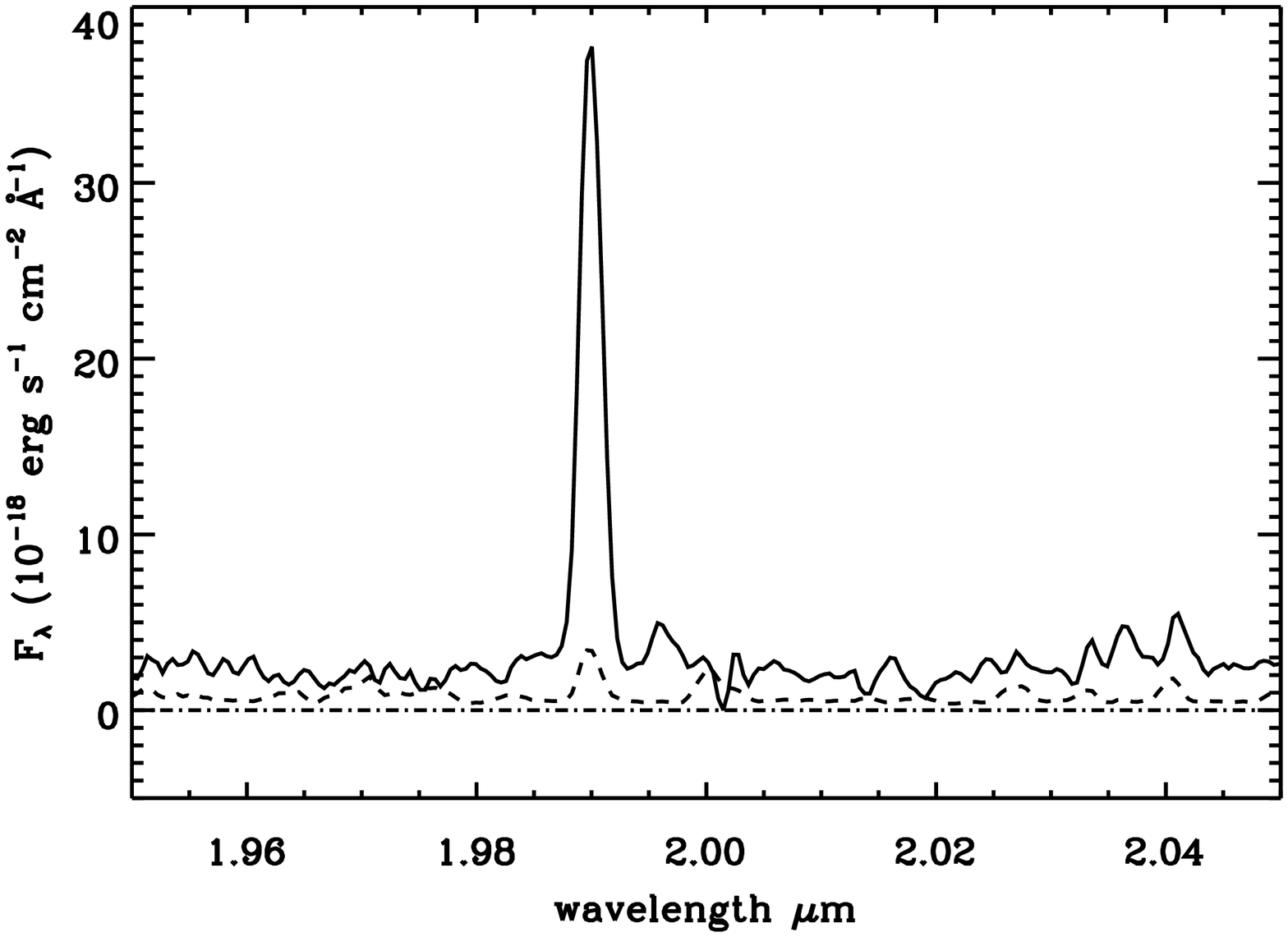}
\caption{One-dimensional $H$+$K$ band spectra for the two components 
of J0900+2234.The upper panel is the spectrum of component A and the lower 
panel is the spectra of component B. The dashed line is $1\sigma$ error
of the spectrum, and the dash dot line represent the zero flux.  \label{spec}}
\end{figure}

To reduce the LUCIFER spectra, we modified an IDL long-slit reduction package 
written by G. D. Becker for NIRSPEC \citep{beck09}. 
In this package, the following sky background 
subtraction procedure is used. First, four types of calibration
files were created: a median-combined normalized
internal flat field to correct the pixel-to-pixel variations in QE,
 a median-combined dark image, and two 
transformation maps created by tracing
bright standard stars and bright sky lines.
These two transformation maps were used to transform
the NIR detector (x, y) coordinates to the slit position and 
wavelength coordinates. Then, the dark image was subtracted from the science frames
which were then divided by the flat field. The background in each science
frame was subtracted using a fit to the science frames 
obtained before and after it 
in two dimensions, based on the transformation maps.
A second-order polynomial function and a $b$-spline function were
used to fit the sky background residual along the slit direction 
and the dispersion direction, respectively.
After the sky background subtraction, the x and y direction distortion of
the science frames were corrected based on the transformation maps
created in the first step.
The one-dimensional spectra were extracted from individual 
two-dimensional spectrum frames and combined together
to give the averaged spectrum shown in Fig.~\ref{spec}. The wavelength calibration 
was derived from observations of Ar lamps and applied to the 
averaged spectrum. 
The spectrum was corrected for telluric features with 
the spectrum of an A\Rmnum{5} star observed at the same average air mass 
as the science exposures. Finally, we derived the flux calibration by 
normalizing the spectrum to the $H$-band magnitude.

\section{Measurements}\label{measurements}

\subsection{NIR Photometry: Discovery of a new Lensed Galaxy}\label{phot}
\begin{figure}
\center
\epsscale{0.8}
\plottwo{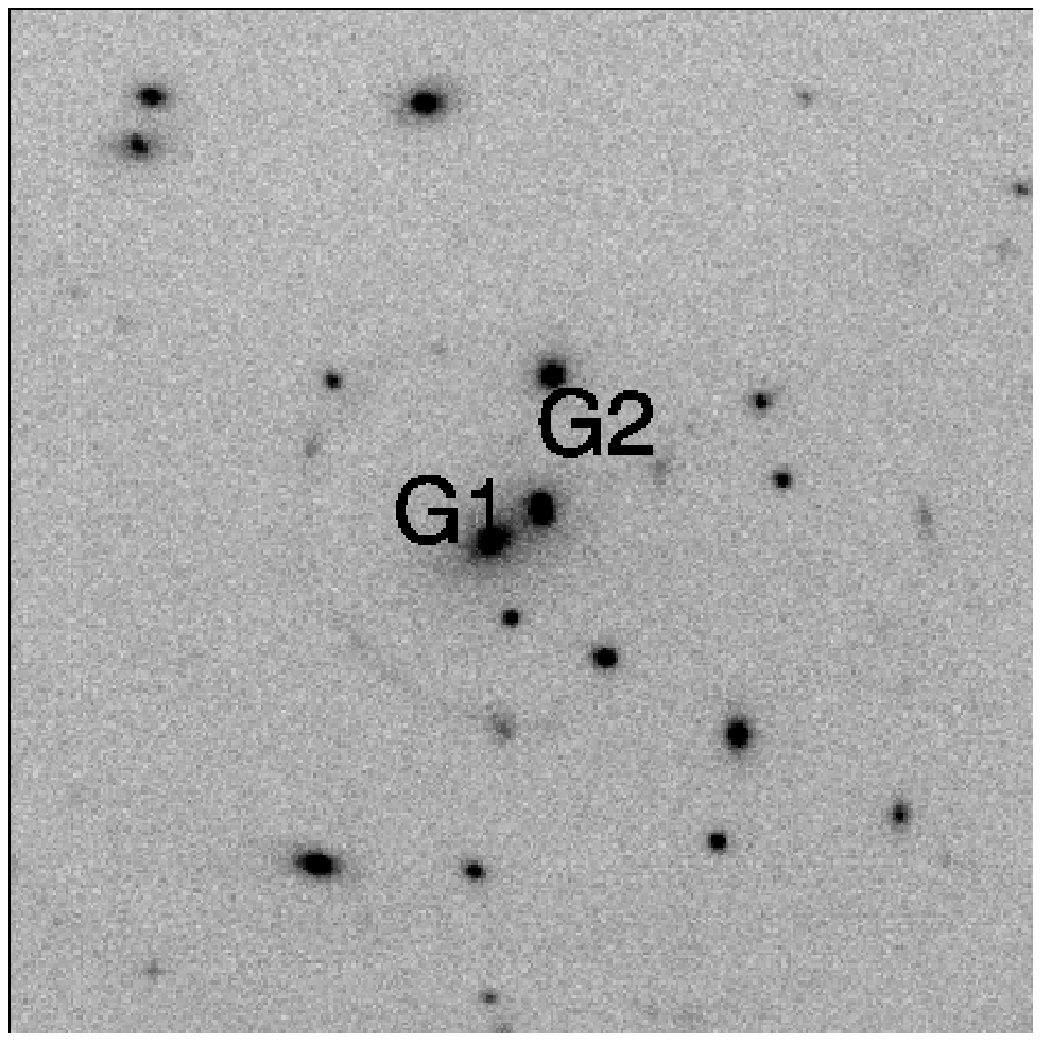}{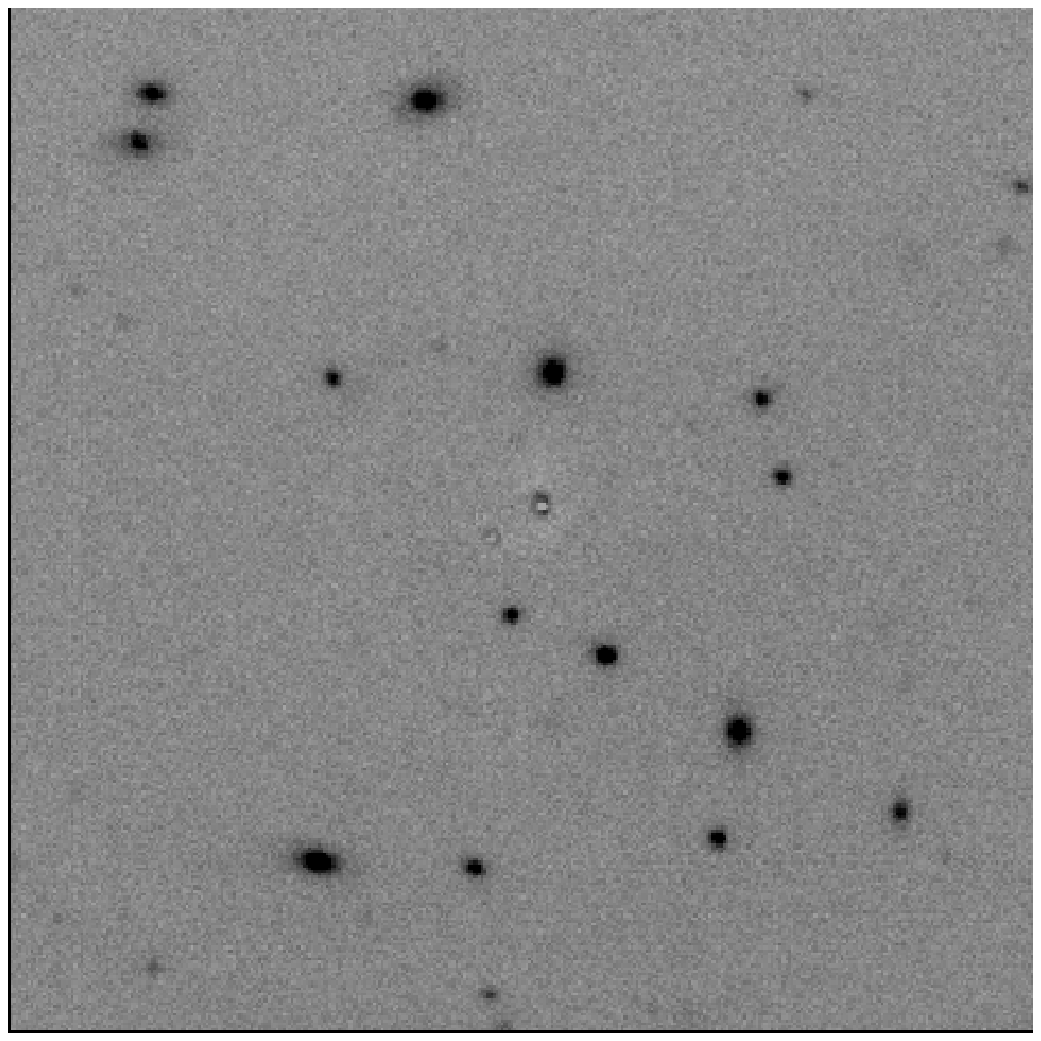}
\caption{The $Ks$-band image and the residual image with all model components fit
by GALFIT subtracted.\label{galfit}}
\end{figure}

The GALFIT program \citep{peng02,peng10} was used to model the central lensing galaxies and lensed 
components. We fit each of the two central massive galaxies with de Vaucouleurs models, and
fit the three lensed knots and two arcs with exponential disk models.
The initial input parameters -- namely, 
 position, total magnitude, axis ratio, and position angle for each component -- were determined using SExtractor \citep{bert96}.
We used the $Ks$ band co-added image to fit the profile parameters (Fig.~\ref{galfit} and Table~\ref{galfit1}), 
and fixed these parameters when
fitting the $J$ and $H$-band images to obtain the magnitude of each component (Table~\ref{galfit2}). 
The values of best-fit reduced $\chi^2$ for $J$, $H$, and $Ks$ bands are 1.064, 1.094, and 1.088, respectively.

The NIR colors of the A and B components are similar, with $J-H \sim 0.1$ and $H-Ks \sim 0.0$, while the colors of
the C and D components appear to be different from those of the A and B components. For D component, 
the reason for this discrepancy may be that the exponential disk profile is 
not a good fit to the observed arc structure. 
For the C component, the follow-up spectroscopy was performed with MMT/Blue Channel spectrograph, 
which shows that the C component is not part of this lens system.

Current data are not yet deep enough to uncover other components of the weak A1 component. 
If confirmed by spectroscopy and deeper imaging, this newly discovered arc would make J0900+2234
a rare example of a double Einstein ring system lensed by galaxy.

 

\subsection{Emission Line Measurements}

We used the IDL MPFIT\footnote{http://cow.physics.wisc.edu/~craigm/idl/idl.html} package to fit the
the emission lines. The results are listed in
Table~\ref{line}. The H$\beta$, [\ion{O}{3}]$\lambda4959$ and [\ion{O}{3}]$\lambda5007$ were fit as
Gaussian functions individually. The positions of H$\alpha$ and [\ion{N}{2}]$\lambda6583$ as well as
[\ion{S}{2}]$\lambda\lambda6717,6732$ are close to each other, so they were fit by two Gaussian 
functions together to deblend these two lines. The observed line wavelength, line flux, and full width
at half maximum (FWHM) were derived from the Gaussian fitting (Table~\ref{line}). Monte Carlo (MC) simulation was 
used to estimate the uncertainties of line flux and FWHM. One thousand artificial spectra were generated
by perturbing the flux of each data point from the true spectrum by a random amount proportional to the $1\sigma$
flux error. We then measured the line flux of each fake spectrum with leaving all the parameters free
for both the single lines and deblened line pairs.
The standard deviations of  the distributions of the line flux and the FWHM were adopted as
the uncertainty of measurements of the line flux and the FWHM (Table~4).
The emission line redshifts of A and B knots are $2.0321\pm0.0009$ and $2.0318\pm0.0005$,
respectively, and are consistent with each other within $\Delta z/(1+z)= 0.0001$.
These results also agree with the mean redshift from the optical spectra, $z=2.0325\pm0.0003$ \citep{dieh09}.

\section{Lensed Model}\label{lensmodel}
The lens modeling is performed using LENSFIT \citep{peng06}.
Briefly, LENSFIT is patterned after, and works like GALFIT \citep{peng02,peng10}.
LENSFIT is designed to specifically deal with situations where the image geometry 
is crowded: it allows one to decompose foreground and background galaxy light profiles, 
and determine the lens deflection model, simultaneously.  
The number of light profile and singular isothermal ellipsoid (SIE) deflection models is unrestricted, 
and the optimization process is done using the Levenberg-Marquardt algorithm in 
Numerical Recipes \citep{press92}.

In J0900+2234, 
the deflection model is potentially quite complicated because 
the system is embedded in a compact cluster environment. 
There are at least two primary deflectors, and potentially 5 in all, 
just within the Einstein ring of the system which has a radius of roughly 7.7 arcseconds.
Due to the low signal-to-noise of the source detection, it is 
difficult to determine an accurate lens model. However, it is clear from 
visual inspection that a single SIE deflector probably 
would not suffice because the geometric center of the lensed arcs 
falls in the gap between the two primary deflectors.
Therefore, the simplest lens deflection model we adopt is that of 
two SIEs held to the position of the light profile of 
the two primary lensing galaxies.
We also hold the SIE axis ratios fixed to the light profile models of the lenses,
but allow the Einstein ring radii and position angle parameters 
to be free in the fit.
We simultaneously optimize the light profiles of all the foreground 
sources and the background galaxy light profiles, in all 14 objects
in $J$-band image.

With the limitation of our lens model and the data S/N in mind, 
we infer that the size of the background source is 0.8-1.0 arcsecond
which corresponds to the physical size of 6.6-8.2~kpc,
in deprojected angular size,
and it has a luminosity of 21.7 to 21.8 mag (AB) in $J$ band,
given a magnification of the A and B components together
is ~4.6-5.0. For the further analysis, we use the
effective radius of 7.4~kpc and the A and B component
magnification of 4.8. The Sersic index is quite high, 
from n=3.5 to n=4. The major uncertainty of the 
lens fitting is caused by the influence of the  bright
foreground source C, which is right on the 
Einstein ring and the source elsewhere has quite low S/N.

\section{Physical Properties}\label{physics}
With LUCIFER we obtained $H$+$K$ (1.40-2.20 $\mu$m) spectra of the A and B components 
of the lensed galaxy J0900+2234 (Fig.~\ref{spec}). The 
H$\beta$, [\ion{O}{3}]$\lambda\lambda$4959,5007, 
H$\alpha$, [\ion{N}{2}]$\lambda$6583, and [\ion{S}{2}]$\lambda\lambda$6717,6732 lines were 
detected. In this section, we discuss
the physical properties which can be derived from these observations.

\subsection{Emission-Line diagnostics}
\begin{figure}
\center
\epsscale{0.6}
\plotone{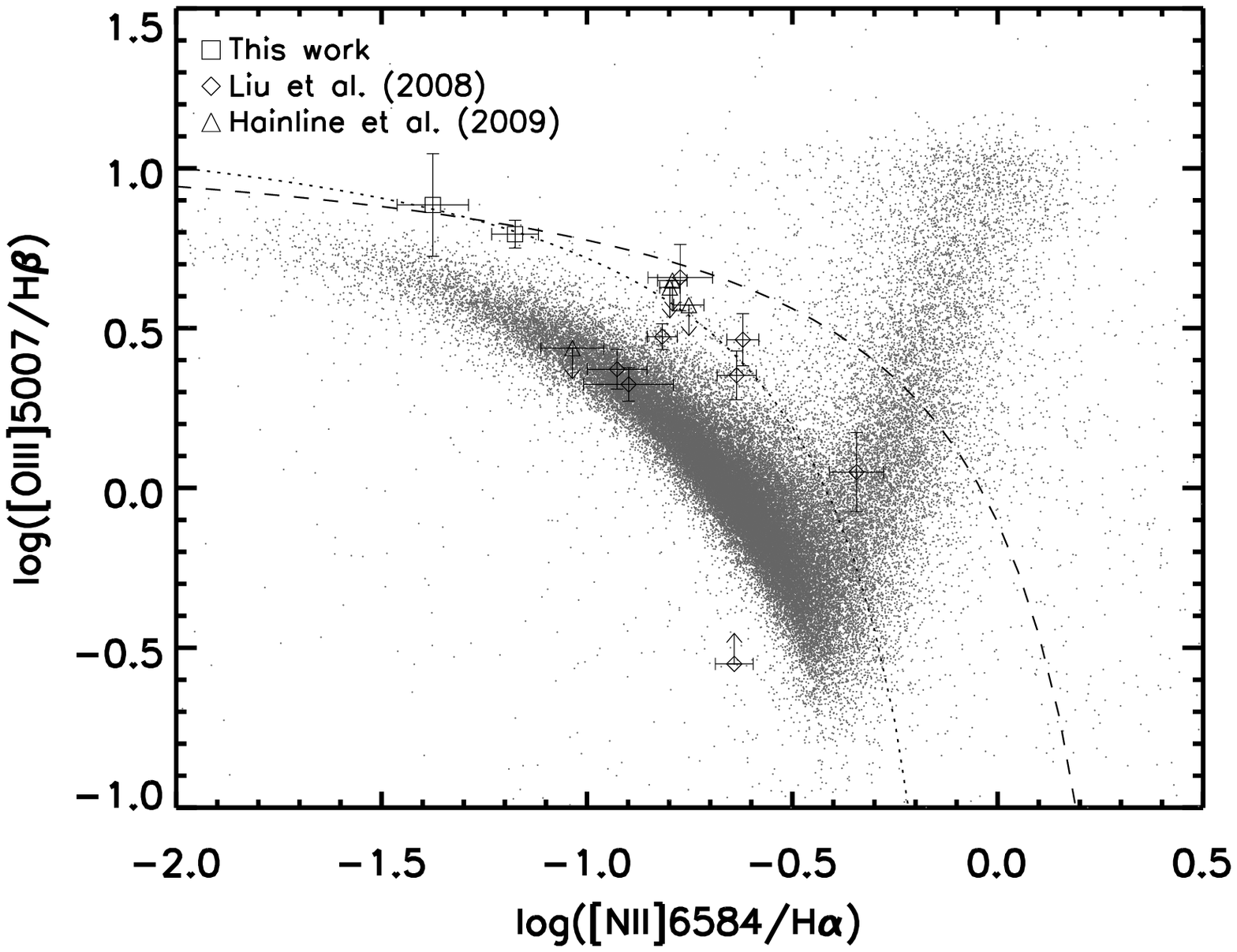}
\caption{The \ion{H}{2} region diagnostic diagram of log(\ion{N}{2}$\lambda6584$/H$\alpha$)
and log(\ion{O}{3}$\lambda5007$/H$\beta$). The open rectangles show the location of 
the A (left) and B (right) components. The open diamonds represent $z\sim1-1.5$ DEEP2
objects \citep{liu08} and the open triangles represent $z\sim2$ lensed star-forming galaxies
\citep{hain09}
The grey points represent SDSS star-forming galaxies and 
AGNs. The dotted line and dashed line are
empirical \citep{kauf03} and theoretical \citep{kewl01} separation of star-forming galaxies and AGNs.
\label{bpt}}. 
\end{figure}

First, it is important to examine the observed values of well-known empirical 
line ratio 
diagnostics to determine whether the emission is from an \ion{H}{2} region, active 
galaxy, or shocked gas. The line ratios of the A and B components on the 
diagram of \citet{bald81} (BPT diagram) are shown in Fig.~\ref{bpt}.
The weak \ion{N}{2} emission rules out the possibility that the emission is from 
the activity of a central active galactic nucleus (AGN).
The location of the A and B components on the BPT diagram are similar to other high-z star forming
galaxies \citep[e.g.,][]{liu08,hain09}. They have relatively higher
[\ion{O}{3}]$\lambda5007$/H$\beta$ values and lower [\ion{N}{2}$\lambda6584$/H$\alpha$
values compared to the local star-forming galaxies from SDSS (Fig.~\ref{bpt}).
This offset could be the result of relatively intensive star formation activity and
high electron density in these high redshift galaxies, compared to the local sample.
Overall, we conclude that the line emission is from gas that is photo-ionized by a hot 
stellar continuum, that is, the emission is from star-formation regions. 

\subsection{Reddening and Extinction}
We estimated the extinction of the lensed galaxy by two methods: (1) 
a fit to the stellar continuum as measured by broad band photometry, and (2) Balmer decrement.
These two methods represent dust extinction to the stellar continuum $E_s(B-V)$ and 
the dust extinction to the nebular gas in the \ion{H}{2} region ($E_g(B-V)$), respectively. 
One expects these two estimates to differ, and indeed some previous studies 
have observed that there is more extinction towards the 
ionized gas than the stellar continuum in local star-forming galaxies \citep{calz00} and high redshift
galaxies \citep[e.g.,][]{fors09}. In contrast, 
\citet{erb06b} did not find any extinction difference between ionized gas and the 
stellar continuum in star-forming galaxies at redshift of $\sim2$. 
For lensed galaxies, the reddening derived from individual images
often differs because of reddening by different amounts of foreground dust, 
since the images traverse different sight-lines through the lensing galaxies.


%
%

With optical photometry alone, one can measure only the UV continuum shape in high redshift 
galaxies, making it 
difficult to distinguish between reddening by dust and age of the stellar population.
By adding the NIR photometry, we can measure the 
Balmer break (3600-3700 \AA) which is sensitive to the 
age of the stellar population, and the 4000 {\AA} break strength only weakly depends on metallicity.
By combining our NIR photometry with the optical, we can fit stellar 
population models and simultaneously determine the age of the galaxy and average extinction.

\begin{figure}
\center
\epsscale{0.8}
\plotone{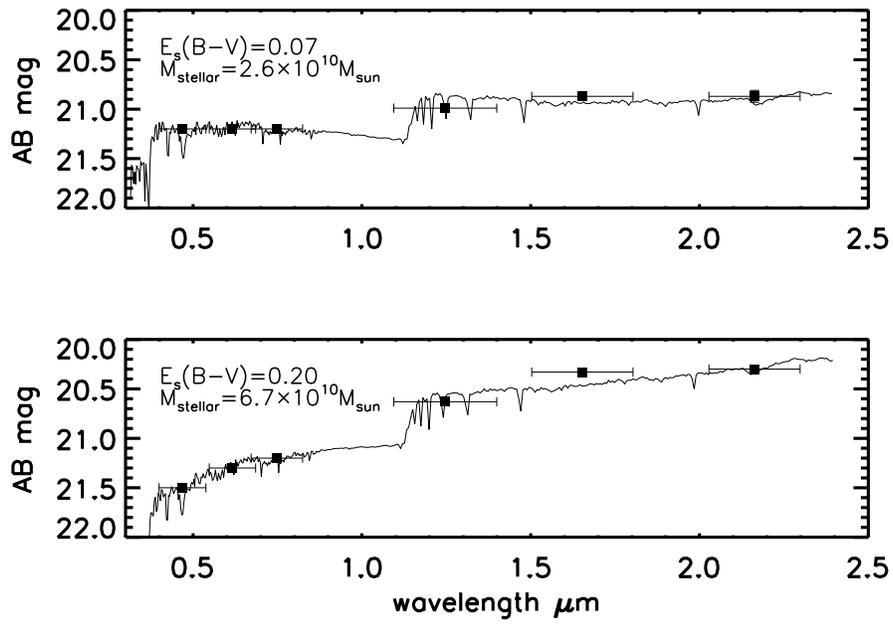}
\caption{$g$, $r$, $i$, $J$, $H$ and $Ks$-band AB magnitudes of the knot A (the upper panel)
and the knot B (the bottom panel) and their best-fit model spectra.The best fit age is 180 Myrs for
both knots and  
the lensed corrected stellar mass is $1.9\times10^{10}M_\sun$. \label{sedfit}}
\end{figure}

Here we used the $J$, $H$, and $Ks$-band (rest-frame optical bands) magnitudes
combining with the $g$, $r$, and $i$-band (rest-frame UV bands)
magnitudes (3$^{\prime\prime}$ from \citet{dieh09} to estimate the $E_{\rm s}(B-V)$ 
and the age of the galaxy. The \citet[][BC03]{bruz03} standard simple stellar population 
(SSP) model was used to build the spectral templates of star-forming galaxies. 
We adopted constant star formation rate (SFR) with the Chabrier initial mass function (IMF) \citet{chab03}, 
and solar metallicity, to generate a series of spectra with different ages and reddening.
Fig.~\ref{sedfit} shows the best-fit spectra with the photometry data of A and B components.
From the fitting, we found that the ages of the components A and B were consistent with each
other, and equal to 180~Myr, with the values of $E_s(B-V)$ equal to 0.07 and 0.20 for the 
components A and B, respectively. Based on the metallicity found in section~\ref{metal},
we also generated the spectra with 0.25 solar metallicity to fit the broad band photometry
data, and found similar results.

With this fit to the stellar population we derive 
the intrinsic stellar mass of
the source galaxy is $1.9\times10^{10}~M_\sun$ after correcting for lensing magnification,
which is smaller than the mean stellar mass of $3.6\times10^{10}~M_\sun$
found in a $z\sim2$ UV-selected galaxy sample \citep{erb06c}. 
Using $Ks$ band magnitude to approach the rest-frame $R$ band magnitude,  
we found the  $M/L_R$ in J0900+2234 is 0.13.
\citet{shap05} found that the mean of  $M/L_R$ is 0.29 with a large scatter from 0.02
to 1.4 (where the values have been divided by 1.8 to convert the Salpeter IMF
to Chabrier IMF) in $z\sim2$ UV-selected star-forming galaxies. Considering
the high scatter, $M/L_R$ in J0900+2234 agrees with other $z\sim2$ star-forming
galaxies. The intrinsic UV ($\sim1700$~\AA) brightness of the J0900+2234 is 22.1 magnitude,
so that the absolute magnitude is M(1700)$_{\rm AB}$ = -22.7, compared to 
L* = -20.20$-$-20.97 at $z\sim2$ \citep[e.g.][]{oesch10,redd08,redd09}. 
Thus, J0900+2234 has L/L* $\approx$ $5-10$, and is an intrinsically 
luminous galaxy. Galaxies with these high luminosity is very rare, and their number density is 
about a few $10^{-6}$~Mpc$^{-3}$~mag$^{-1}$ base on the luminosity function from \citet{redd08} and
\citet{redd09}.  The high intrinsic UV luminsity indicates a high SFR 
(see details in Section~\ref{SFR}) in this galaxy.

We estimated $E_g(B-V)$ of J0900+2234 using the flux ratio of H$\alpha$ and H$\beta$ lines.
Under Case B \ion{H}{1} recombination at a temperature T=10,000~K and electron density n
$\le $100~cm$^{-3}$ \citep{zari94}, the intrinsic value of H$\alpha$/H$\beta$ is 2.86
\citep{oste06}. Using the extinction law proposed by \citet{calz00}, we found the 
values of $E_g(B-V)$ were $0.84\pm0.31$ and $0.59\pm0.08$ for the A and B components.
For both A and B components, the values of $E_g(B-V)$ are significantly larger than 
those of $E_s(B-V)$, in agreement with the results of \citet{calz00} and \citet{fors09}.
The stars and line-emitting gas are not co-spatial, and the line-emitting regions are 
dustier than the galaxy as a whole. 
Note that the low S/N of the H$\beta$ line in knot a makes the uncertainty of the 
$E_g(B-V)$ for component A large. Therefore, we use the value of $E_g(B-V)=0.59\pm0.08$ of knot B
for further analysis.

\subsection{Star Formation Rate}\label{SFR}

We estimated the star-formation rate (SFR) in J0900+2234 using the
luminosity of H$\alpha$ \citep[$L_{\rm H\alpha}$,][]{kenn98}.
The relation between the SFR and $L_{\rm H\alpha}$ is 
\begin{equation}
{\rm SFR}(M_\sun \rm yr^{-1})=7.9\times10^{-42}\frac{L_{\rm H\alpha}}{\rm ergs~s^{-1}}
\end{equation}
The H$\alpha$ fluxes of the A and B components are $37.01\pm0.79 \times10^{-17}$~ergs~s$^{-1}$~cm$^{-2}$
and $84.64\pm3.21 \times10^{-17}$~ergs~s$^{-1}$~cm$^{-2}$. We used $E_g(B-V)=0.59\pm0.08$ to correct the 
fluxes for extinction and found the extinction-corrected fluxes are $(225.5\pm55.5)\times10^{-17}$~ergs~s$^{-1}$~cm$^{-2}$
and $(516.0\pm128.0)\times10^{-17}$~ergs~s$^{-1}$~cm$^{-2}$. The 
extinction-corrected luminosities are therefore $(15.4\pm3.8)\times10^{43}$~ergs~s$^{-1}$ 
and $(6.75\pm1.66)\times10^{43}$~ergs~s$^{-1}$.
The lensed SFRs of components A and B are $533\pm131 ~M_\sun~\rm yr^{-1}$ and $1220\pm303~M_\sun~\rm yr^{-1}$.
We added the components A and B SFRs together, corrected for the magnification, and found that
the intrinsic SFR of the lensed galaxy was $365\pm69~M_\sun~\rm yr^{-1}$. Note that if we use the
Chabrier IMF, the SFR derived from the $L_{H_\alpha}$ will be $203\pm38~M_\sun~\rm yr^{-1}$. 
We also estimate the SFR from the UV luminosity \citep{kenn98}  which is derived from the average of 
$g$ and $r$ band flux. We found that the SFR is $276~M_\sun~\rm yr$ for Shalpeter IMF and 
$153~M_\sun~\rm yr$ for Chabrier IMF by correcting the dust extinction, $E_s(B-V) = 0.07 (0.20)$ for A (B) components.
The SFR based on the dust-corrected H$\alpha$ emission is consistent with that from the dust-corrected 
UV emission. If we use $E_s(B-V)$= 0.20 to correct the extinction of H$\alpha$ flux, the 
H$\alpha$-derived SFR of $110~M_\sun~\rm yr^{-1}$ will be significant lower than UV-derived 
SFR. This supports the result that the $E_g(B-V)$ is larger than $E_s(B-V)$.   
The average of the SFR for $z\sim2$ star-forming galaxies is 
\textlangle$\rm SFR_{\rm H\alpha}$\textrangle$=31~M_\sun~\rm yr^{-1}$ and
\textlangle$\rm SFR_{\rm UV}$\textrangle$=29~M_\sun~\rm yr^{-1}$\citep{erb06c}.
Despite the uncertainties in interpreting the SFR measurements, 
we conclude that the SFR of J0900+2234 is an order of magnitude higher than 
typical star-forming galaxies at $z\sim2$.

\subsection{Oxygen Abundance}\label{metal}
Emission lines from \ion{H}{2} regions can be used to measure metallicity. 
The oxygen abundance was calculated
using the following indicators: the 
N2 index (N2$\equiv\log(F_{\rm{[N II]}\lambda6584}/F_{\rm H\alpha})$) and 
the O3N2 index (O3N2$\equiv\log\{(F_{\rm [O III]
\lambda5007}/F_{\rm H\beta})/(F_{\rm [N II]\lambda6584}/F_{\rm H\alpha)}\}$), which have been well calibrated
using nearby extragalactic \ion{H}{2} regions to measure O/H \citep{pett04}.
The advantage of using the N2 and O3N2 indices as metal abundance estimators
 is that H$\alpha$ and the [\ion{N}{2}]$\lambda6584$ pair, as well
as the H$\beta$ and [\ion{O}{3}]$\lambda5007$ pair, are close in wavelength, making 
these two estimators relatively insensitive to extinction.

The N2 index is sensitive to the oxygen abundance \citet{stor94}, and 
was further calibrated by \citet{raim00} and 
\citet{deni02}. Here we use the best linear fit between
the N2 and ($12+\log(\rm{O/H})$) from \citet{pett04}, 
\begin{equation}
12 + \log(\rm{O/H}) = 8.90 + 0.57 \times \rm{N2},
\end{equation}
where N2 is in the range from -2.5 to -0.3, and the 1$\sigma$ error of the measurements of 
$\log(\rm{O/H})$ is 0.18. We find the values of $12+\log(\rm{O/H})$
equal to $8.12\pm0.21$ and $8.23\pm0.19$ for components A and B, respectively, which are about
$0.27\pm0.13$ to $0.35\pm0.15$ solar abundance (for the sun: $12+\log(\rm{O/H_{\sun}})$ = 8.69 \citep{aspl09}).

The O3N2 index was first introduced by \citet{allo79}, and was well calibrated by \citet{pett04} using
137 extragalactic \ion{H}{2} regions. The relation between the metallicity ($12+\log(\rm{O/H})$)
and the O3N2 index is:
\begin{equation} 
12 + \log(\rm{O/H}) = 8.73 - 0.32 \times \rm{O3N2},
\end{equation} 
where O3N2 is in the range from -1 to 1.9, and the 1$\sigma$ error of the measurements of
$\log(\rm{O/H})$ is 0.14, which indicates a less scattering than the N2 index.
We found the values of $12+\log\rm{(O/H)}$ of components
A and B were $8.00\pm0.16$ and $8.09\pm0.15$, which are $0.21\pm0.08$ and
$0.25\pm0.08$ solar abundance. 

\begin{figure}
\center
\epsscale{0.6}
\plotone{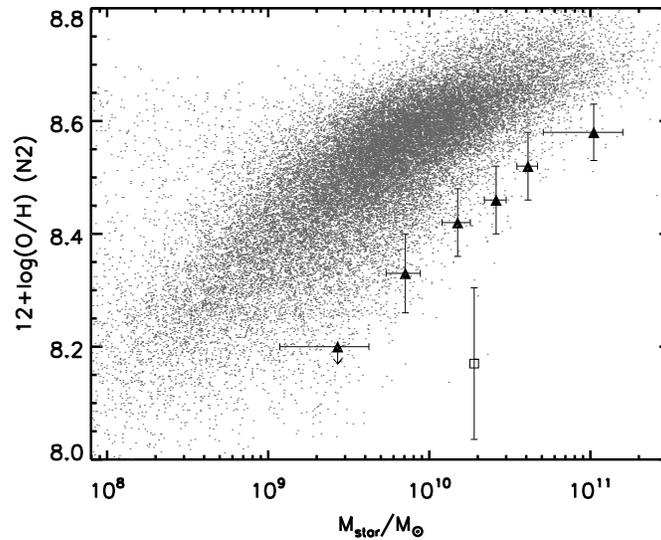}
\caption{The stellar mass and metallicity relation for the SDSS galaxies 
(grey points) and the galaxies at $z\sim2$ \citep[filled triangle][]{erb06a}.
The open rectangle represents the lensed galaxy J0900+2234, which s significant
lower than the relation in the galaxies at $z\sim2$.\label{massmetall}}
\end{figure}

Both the N2 and O3N2 indices indicate that J0900+2234 has a 
low oxygen abundance compared with other $z\sim2$ star-forming galaxies. 
Fig.~\ref{massmetall} shows the stellar mass and metallicity relation in the local 
SDSS starburst galaxies and $z\sim2$ star-forming galaxies \citep{erb06a}. The metallicity is
derived from N2 index. We averaged the value $12 + \log(\rm{O/H})$ of the 
A and B components and the stellar mass is from the broad band fitting result.
Our data point is significantly lower than the mass and metallicity relation
in star-forming galaxies at $z\sim2$\citep{erb06a}. \citet{man10} proposed
a more general fundamental relation between stellar mass, SFR and 
metallicity derived with SDSS galaxies, 
in which the metallicity decreases with the increase of SFR. 
We used the relation derived in \citet{man10} (equation 2) to obtain
the predicted metallicity  from SFR and stellar mass and found the 
value is 8.45. Although our measured metallicity is $\sim0.27$ dex lower than
the predicted value, this dispersion is consistent with the results of 
\citet{man10} that the distant galaxies show 0.2-0.3 dex dispersions
for the fundamental metallicity relation.  


\subsection{Electron Density}
The flux ratio of [\ion{S}{2}]$\lambda6717$ and [\ion{S}{2}]$\lambda6732$ was 
used to estimate the electron 
density in the \ion{H}{2} region of J0900+2234. The IRAF task
$stdas.analysis.nebular.temden$ was used to compute the density. 
We assumed that the temperature in these regions were 10,000~K. 
We found the values of $F_{\rm[S II]\lambda6717}$/$F_{\rm[S II]\lambda6732}$ 
were $0.88\pm0.24$ and $0.84\pm0.31$ for the components A and B. The large
error bar is due to the low S/N ($<5$) of these [\ion{S}{2}] lines in both
apertures. The values of the ratio yielded a range of electron number 
density $1029^{+3333}_{-669}$~cm$^{-3}$ and
$1166^{+7020}_{-855}$~cm$^{-3}$, respectively for these two components. Nonetheless,
these densities are much higher than those typical of local \ion{H}{2} regions 
$n_e\sim$100~cm$^{-3}$ \citep[e.g.,][]{zari94}. The electron density of J0900+2234
is similar to that in the lensed galaxies, the Cosmic Horseshoe and the Clone, which is
also derived from the ratio of  [\ion{S}{2}] double lines \citep{hain09}.
The high electron density was also found in 
the Cosmic Horseshoe with 5 000$-$25 000~cm$^{-3}$
derived from the ratio of \ion{C}{3}]$\lambda\lambda$1906, 1908 doublet \citep{quid09}. 
The high electron
density implies the compact size of the \ion{H}{2} regions in 
the high-z galaxies. If these
high-z \ion{H}{2} regions follow the similar electron density-size relation
found in the local galaxies \citep{kim01}, their sizes should be 
less than 1~pc.


\subsection{Virial Mass and Gas Mass}
The width of the emission lines can be used to probe the dynamics and the total mass of the
parent galaxy. By fitting the emission lines (e.g., H$\alpha$, [\ion{O}{3}]) as Gaussian profiles, we found
that the FWHMs of H$\alpha$ of A and B knots were 19.0~{\AA} and 21.8~{\AA}. These observed 
FWHMs were
corrected by subtracting the instrument resolution ($\sim16~$\AA) in quadrature. The corrected FWHMs
were converted to the 1-D velocity dispersion with $\sigma$ = FWHM/$2.355\times c/\lambda$. 
The values of $\sigma$ for the A and B knots are $66\pm5$~km~s$^{-1}$ and $95\pm6$~km~s$^{-1}$ respectively. 
We averaged the $\sigma$ components of A and B to estimate the velocity dispersion of J0900+2234,
which is $81\pm4$~km~s$^{-1}$.  Our velocity dispersion is about 30\% lower than that in a sample of 
$z\sim2$ star-forming galaxies\citep{erb03}, who found a mean velocity dispersion of $\sim110$~km~s$^{-1}$. 

The mass of the galaxies can be estimated by assuming a simplified case of 
a uniform sphere \citep{pett01}
\begin{equation}
M_{vir} = \frac{5\sigma^2r_{1/2}}{G}
\end{equation}
\begin{equation}
M_{vir} = 1.2\times10^{10}M_{\sun}\left(\frac{\sigma}{100~\rm{km}~\rm{s}^{-1}}\right)^2\frac{r_{1/2}}{\rm{kpc}}
\end{equation}
where $G$ is the gravitational constant and $r_{1/2}$ is the half-light radius, which is $7.4\pm0.8$~kpc from 
the lens model. We find that the dynamical mass of J0900+2234 is $(5.8\pm0.9)\times10^{10}~M_\sun$, an 
upper limit since the line emitting gas may not be in Virial equilibrium with the gravitational 
potential of the galaxy.  

Using the global Kennicutt-Schmidt law \citep{kenn98}, we convert the surface gas density 
($\Sigma_{gas}$) with the SFR derived from H$\alpha$:
\begin{equation}
\Sigma_{gas} = 1.6\times10^{-27}\left(\frac{\Sigma_{H\alpha}}{erg^{-1}kpc^{-2}}\right)^{0.71}~M_{\sun}~pc^{-2}
\end{equation}
where $\Sigma_{H\alpha}\simeq L_{H\alpha}/r_e^2$ is the surface density of H$\alpha$  
luminosity \citep{erb06b,fink09}. Then the gas mass, $M_{gas}$, can be derived from   $\Sigma_{H\alpha}\times r_e^2$,
which is $5.1\pm1.1\times10^{10}~M_{\sun}$.  And the gas fraction, $f_{gas}=M_{gas}/(M_{gas}+M_{stellar})$, is $0.74\pm0.19$.
\citet{erb06a} found the gas fraction increases when the stellar mass decreases in the UV-selected 
star forming galaxies at $z\sim2$.   
The gas fraction in J0900+2234 is higher than the mean gas fraction ($0.48\pm0.19$) 
of  \citet{erb06a} galaxies in the similar stellar mass bin ($1.5\pm0.3\times10^{10}~M_{\sun}$), but still 
$1\sigma$ consistent. The total baryonic mass $M_{gas}+M_{stelar}$ is $6.9\pm1.1\times10^{10}~M_{\sun}$,
which is in the agreement with the dynamic mass.


\subsection{Measurement Uncertainty}
The main uncertainty of the measurement of the extinction and abundance is from the uncertainties
in the measurements of the weak emission lines (e.g. H$\beta$ and \ion{S}{2}) in the observed NIR.
There are many prominent sky emission lines in the NIR, especially in the $H$ band,
which can contaminate our line measurements, especially for the weak emission lines.
In the LUCIFER spectra of J0900+2234, the H$\beta$ sits near the blue cut of the H band,
where there are several moderate strong sky emission lines and the instrument efficiency, 
which contributes large flux errors (S/N$<5$) and underestimate/overestimate the fluxes.  
The higher $E_g(B-V)$ value of the A (fainter) component and  
systematically lower metallicities estimated from O3N2 index compared to those from N2 index
may imply the flux of H$\beta$ is underestimated. 

The systematic uncertainty of SFR, associated with the absolute flux of the H$\alpha$, 
is mainly due to the uncertainties in the absolute calibration of the NIR spectra. 
We calibrated the spectra with an AV star, and 
scaled the spectral flux to the H band magnitude to correct the light loss in the 
spectrograph slit. The scale factors are $\sim1.6$ ($\sim1.8$) for the A (B) knots, which is 
a reasonable value for light loss. Converting the lensed H$\alpha$ luminosity to the intrinsic 
H$\alpha$ luminosity is another uncertainty for SFR estimation. The uncertainty of the 
lens model fitting is $20-30\%$. The uncertainty also affect the estimation of the 
stellar mass and dynamical mass, which depend on the results of the lens model.
The H$\alpha$ and SFR relation \citep{kenn98} 
is derived from stellar synthesis models with solar metallicity. The SFR in J0900+2234
could be overestimated, given the low metallicity ($\sim0.25$ $Z_{\sun}$) in this
galaxy.

\section{Conclusions}\label{conclusion}
We present LBT/LUCIFER1 NIR (rest-frame optical) imaging and spectroscopy 
of the lensed galaxy J0900+2234 ($z=2.03$). The lensed components
A and B were placed in the slit to obtain NIR spectra covered
from 1.40 to 2.20$~\mu$m. The detailed physical properties of the lensed 
star-forming galaxy were studied using the rest-frame optical spectra (Table~\ref{phys}).
We summarize the main results as follows. 

\begin{itemize}
 
\item The new imaging was used to construct a lensing model. The magnification factor for
the fluxes of the lensed galaxies (A plus B) is estimated to be 4.8.

\item A new lensed arc A1 was discovered by the deep NIR $J$, $H$, and $Ks$-band images.
The colors and position imply that this arc does not have the same source galaxy as the 
other four components discovered before, suggesting the presence of a 
rare double Einstein ring. Follow-up spectroscopy shows that the C component does
not have the same source galaxy as A and B.

\item We fit the optical and NIR broad band photometry to theoretical stellar population 
spectral templates (BC03),
and found the galaxy age to be 180~Myr, $E_s(B-V) = 0.07$ ($E_s(B-V)=0.20$) for the A (B) 
component. The stellar mass is 1.9$\times10^{10}~M_\sun$, and the intrinsic luminosity of the galaxy is 
L/L* $\approx$ $5-10$.

\item Using the flux ratio of H$\alpha$ and H$\beta$, we found that the 
emission line gas suffers an extinction of 
$E_g(B-V)$ is $0.84\pm0.31$ ($0.59\pm0.08$) for the A (B) component, which is much higher 
than $E_s(B-V)$. This result implies that there is more extinction towards the ionized gas than the 
stellar continuum. 

\item The oxygen abundance was estimated using the N2 and O3N2 indices. 
We found that the metallicity in this star forming galaxy is only $21-35\%$ solar,
which is somewhat lower than typical star-forming galaxies at $z\sim2$. 
The electron number density, $n_e$, was measured using the flux ratio of [\ion{S}{2}]$\lambda6717$
and [\ion{S}{2}]$\lambda6732$. The $n_e$ is in the range of $1029^{+3333}_{-669}$~cm$^{-3}$ ($1166^{+7020}_{-855}$ ~cm$^{-3}$)
for the A (B) component. These values are consistent with other studies of high-z galaxies
\citep[e.g.,][]{hain09,quid09}, but much larger than that in local \ion{H}{2} regions.

\item
The extinction-corrected luminosity of H$\alpha$ emission line was used to estimate the SFR, and the
intrinsic SFR of J0900+2234 is $365\pm69~M_\sun~\rm yr^{-1}$ which is one magnitude higher
than a typical $z\sim2.0$ UV-selected star-forming galaxy.
We estimate the Virial mass of J0900+2234 from the FWHM of the H$\alpha$ emission 
line and find a dynamical mass of $(5.8\pm0.9)\times10^{10}~M_\sun$. The gas mass of the 
galaxy is $(5.1\pm1.1)\times10^{10}~M_\sun$ estimated from the H$\alpha$ flux by assuming the global 
Kennicutt-Schmidt law.

\end{itemize}

Comparison to the local star-forming galaxies, the J0900+2234 (z=2.03)
features as much higher electron density. The high electron density and 
high ionization state \citep[e.g.,][]{hain09}
probably cause the 
offset between the \ion{H}{2}
regions in the high-z galaxies and those in the local galaxies in the BPT diagram. 
This star-forming galaxy also has a much higher SFR, and a relatively lower age, stellar mass, and 
metallicity than those in the typical high-z star-forming galaxies.

The commissioning of LUCIFER1 on the 8.4m LBT left primary mirror (SX)
 has allowed a detailed study of the 
$z=2$ galaxy J0900+2234 in relatively short exposure times. We look forward to
future NIR observations of 
other lensed galaxies, as well as field galaxies which are not lensed, which will give a 
similarly richly detailed picture of galaxies in this important stage of galaxy evolution. 

\acknowledgments
We thank the anonymous referee for informative comments which improved the
manuscript. 
We are indebted to the LBTO staff and LUCIFER team for their great support during
 the observing runs. 
We thank G. D. Becker and X. Liu for discussions about NIR data reduction, and for 
providing their IDL package for our use.
FB, XF, and IDM acknowledge support from a Packard Fellowship for Science and 
Engineering and NSF grant AST 08-06861. 
{\it Facilities:} \facility{LBT}




\clearpage

\begin{table}
\begin{center}
\caption{LUCIFER Observation Log of J0900+2234.\label{log}}
\begin{tabular}{ccccccc}
\tableline\tableline
Date(UT)&mode &Camera &Filter & Slit PA (degree) & Exp. Time(s)\\
\tableline
Jan. 06 2010&imaging &N3.15 & $J$ & - &  $10\times60$ \\
           &&&                $H$ & - &  $10\times60$ \\
           &&&                $Ks$ & - &  $10\times60$ \\
Feb. 13 2010& spectroscopy&N1.8 &$H$+$K$& 86.57 & $24\times300$ \\
\tableline
\end{tabular}
\end{center}
\end{table}
\begin{table}
\begin{center}
\caption{The best-fit profile parameters for the lens galaxies and lensed components.\label{galfit1}}
\begin{tabular}{ccccccc}
\tableline\tableline
Component & Model & Effective Radius ($R_e$($^{\prime\prime}$))  & Axis ratios ($b/a$) & 
Position angle ($\theta$($^{\circ}$))\\
\tableline
G1 & de Vaucouleurs & $4.19\pm0.08$&$0.70\pm0.01$&$-60.63\pm1.00$ \\
G2 & de Vaucouleurs & $1.52\pm0.03$&$0.73\pm0.01$&$7.16\pm1.47$ \\
A  & exponential disk &$0.45\pm0.04$&$0.36\pm0.03$&$-17.84\pm3.22$\\
B  & exponential disk &$0.55\pm0.04$&$0.58\pm0.04$&$-15.70\pm4.93$\\
V  & exponential disk &$0.45\pm0.02$&$0.71\pm0.03$&$48.90\pm5.48$ \\
D & exponential disk &$2.39\pm0.27$&$0.11\pm0.01$&$46.52\pm0.95$ \\
A1 & exponential disk &$0.63\pm0.05$&$0.34\pm0.02$&$12.40\pm2.38$ \\
\tableline
\end{tabular}
\end{center}
\end{table}

\begin{table}
\begin{center}
\caption{Photometry of the LUCIFER images from GALFIT.\label{galfit2}}
\begin{tabular}{ccccc}
\tableline\tableline
Component & $J_{AB}$ & $H_{AB}$ & $Ks_{AB}$ \\
G1 &$17.34\pm0.01$&$16.98\pm0.01$&$16.68\pm0.03$\\
G2 &$18.43\pm0.01$&$18.06\pm0.01$&$17.69\pm0.02$\\
A  &$20.99\pm0.03$&$20.87\pm0.04$&$20.87\pm0.06$\\
B  &$20.63\pm0.02$&$20.33\pm0.04$&$20.30\pm0.04$\\
C  &$20.45\pm0.03$&$20.44\pm0.04$&$19.97\pm0.03$\\
D  &$20.79\pm0.07$&$21.08\pm0.18$&$20.23\pm0.09$\\
A1 &$21.81\pm0.09$&$21.26\pm0.11$&$20.47\pm0.04$\\
\tableline
\end{tabular}
\end{center}
\end{table}

\begin{table}
\begin{center}
\caption{Results of emission line measurements.\label{line}}
\begin{tabular}{cccccc}
\tableline\tableline
Components&Lines & wavelength & flux & FWHM \\
&&\AA&$10^{-17}{\rm~ergs~s^{-1}~cm^{-2}}$&\AA \\
\tableline
A & H$\beta$ $\lambda$4863 &14738.9  &$4.85\pm1.78$ & $18.2\pm8.2$\\
B & H$\beta$ $\lambda$4863&14738.1  &$15.08\pm1.36$ & $18.7\pm1.4$\\
A & [\ion{O}{3}] $\lambda$4959&15043.0&$14.15\pm2.99$ & $22.5\pm8.1$ \\ 
B & [\ion{O}{3}] $\lambda$4959&15037.4& $29.98\pm 3.93$ & $23.1\pm2.6$ \\
A & [\ion{O}{3}] $\lambda$5007&15184.6&$37.24\pm1.78$ & $22.4\pm1.1$\\
B & [\ion{O}{3}] $\lambda$5007&15182.7&$93.81\pm4.07$ & $26.4\pm1.1$ &\\
A &  H$\alpha$ $\lambda$6563& 19902.8 &$37.01\pm0.79$ & $19.0\pm0.4$ \\
B &  H$\alpha$ $\lambda$6563& 19899.5 &$84.68\pm3.21$ & $21.8\pm0.6$\\
A & [\ion{N}{2}] $\lambda$6584&19965.1&$1.56\pm0.35$ & $14.6\pm4.3$\\
B & [\ion{N}{2}] $\lambda$6584&19961.9&$5.66\pm0.88$ & $22.3\pm3.8$\\
A & [\ion{S}{2}] $\lambda$6717&20367.5&$3.56\pm0.50$ & $22.0\pm3.9$\\
B & [\ion{S}{2}] $\lambda$6717&20365.1&$5.11\pm1.01$ & $19.9\pm5.6$\\
A & [\ion{S}{2}] $\lambda$6732&20405.6&$4.06\pm0.98$ & $17.3\pm5.3$\\
B & [\ion{S}{2}] $\lambda$6732&20410.4&$6.06\pm1.87$ & $20.6\pm8.6$\\

\tableline
\end{tabular}
\end{center}
\end{table}

\begin{table}
\begin{center}
\caption{Summary of the physical properties of J0900+2234.\label{phys}}
\begin{tabular}{cccccccccccc}
\tableline\tableline
Knots&age & $E_s(B-V)$ & $E_g(B-V)$ &$Z_{\rm N2}$&$Z_{\rm O3N2}$&$n_e$&
SFR$_{\rm H\alpha}$\tablenotemark{a}&SFR$_{\rm UV}$\tablenotemark{a}&$\log(M_{\rm stellar})\tablenotemark{a}$&$\log(M_{gas})$\tablenotemark{a}&
$\log(M_{vir})$\tablenotemark{a}\\
&Myr & & &$Z_{\sun}$ & $Z_{\sun}$ & cm$^{-3}$ & $M_{\sun}$~yr$^{-1}$ &$M_{\sun}$~yr$^{-1}$ &$M_{\sun}$ & $M_{\sun}$ & $M_{\sun}$ \\
A & 180 & 0.07 & $0.84\pm0.31$ & $0.27\pm0.13$ & $0.21\pm0.08$ & $1029^{+3333}_{-669}$ &
$365\pm69$& $203\pm38$ &10.28 & 10.71&10.76 \\
B & 180 & 0.20 & $0.59\pm0.08$ & $0.35\pm0.15$ & $0.26\pm0.09$ & $1166^{+7020}_{-855}$ &-&-& -& -&-\\

\tableline
\end{tabular}
\tablenotetext{a}{This value is derived from average of A and B components.}
\end{center}
\end{table}


\begin{thebibliography}{}
\bibitem[Ageorges et al. in preparation(2010)]{Ager10}
Ageorges, N. et al.. ``LUCIFER1 commissioning
at the LBT'',  SPIE Proc. Vol. 7735; McLean, Ramsay, Takami (Eds.); in preparation
\bibitem[Allam et al.(2007)]{alla07}
Allam, S. S.; Tucker, D. L., Lin, H., Diehl, H. T., A., J., Buckley-Geer, E. J., Frieman, J. A.
2007, \apj, 662, 51
\bibitem[Alloin et al.(1979)]{allo79}Alloin, D., Collin-Souffrin, S., \& Vigroux, L. 1979, \aa, 78, 200
\bibitem[Asplund et al.(2009)]{aspl09}Asplund, M. Grevesse, N., Sauval, A., \& Scott, P. 2009, \araa, 47, 481
\bibitem[Baldwin, Phillips, \& Terlevich(1981)]{bald81}Baldwin, J. A., Phillips, M. M., \& Terlevich, R., 1981, \pasp, 93, 5
\bibitem[Bechtold et al.(1997)]{bechtold97} Bechtold, J., Yee, 
H.~K.~C., Elston, R., \& Ellingson, E.\ 1997, \apjl, 477, L29
\bibitem[Becker et al.(2009)]{beck09}Becker, G. D., Rauch, M., \& Sargent, W. L. W. 2009, \apj, 698, 1010
\bibitem[Belokurov et al.(2009)]{belo09}Belokurov, V., Evans, N. W., Hewett, P. C., Moiseev, A., McMahon, R. G.,
 Sanchez, S. F., King, L. J. 2009, \mnras, 392, 104
\bibitem[Belokurov et al.(2007)]{belo07}Belokurov, V., Evans, N. W.,
 Moiseev, A., King, L. J., Hewett, P. C., Pettini, M., Wyrzykowski, L., 
McMahon, R. G., Smith, M. C., Gilmore, G., Sanchez, S. F., Udalski, A., Koposov, S., Zucker, D. B., \& Walcher, C. J.
 2007, \apj, 671, 9
\bibitem[Bertin(2006)]{bert06}Bertin, E. 2006, ASPC, 351, 2006
\bibitem[Bertin \& Arnouts(1996)]{bert96}Bertin, E., \& Arnouts, S. 1996, \aaps, 117, 393
\bibitem[Bertin et al.(2002)]{bert02}Bertin, E., Mellier, Y., Radovich, M., Missonnier, G., Didelon, P., \& Morin, B.
 2002, ASPC, 281, 228 
\bibitem[Bruzual \& Charlot(2003)]{bruz03}Bruzual, G. \& Charlot, S. 2003, \mnras, 344, 1000
\bibitem[Calzetti et al.(2000)]{calz00}Calzetti, D., Armus, L., Bohlin, R. C., Kinney, A. L.,
Koornneef, J., \& Storchi-Bergmann, T. 2000, \apj, 533, 682
\bibitem[Chabrier(2003)]{chab03}Chabrier, G. 2003, PASP, 115, 763
\bibitem[Denicol\'{o} et al.(2002)]{deni02}Denicol\'{o}, G., Terlevich, E., \& Terlevich, E., 2002, \mnras, 330, 69
\bibitem[Diehl et al.(2009)]{dieh09}Diehl, H. T., Allam, S. S., Annis, J., Buckley-Geer, E. J., Frieman, J. A., Kubik, D., 
Kubo, J. M., Lin, H., Tucker, D., \& West, A. 2009, \apj, 707, 686
\bibitem[Dickinson et al.(2003)]{dickinson03}Dickinson, M., Papovich, C., Ferguson, H.C., Budavari, T., 2003, \apj, 587, 25
\bibitem[Erb et al.(2003)]{erb03}Erb, D. K., Shapley, A. E., Steidel, C. C., Pettini, M., 
Adelberger, K. L., Hunt, M. P., Moorwood, A. F. M., \& Cuby, J. 2003, \apj, 591, 101
\bibitem[Erb et al.(2006a)]{erb06a}Erb, D. K., Shapley, A. E., Pettini, M., Steidel, C. C., Reddy, N. A., 
\& Adelberger, K. L. 2006a, \apj, 644, 813
\bibitem[Erb et al.(2006b)]{erb06b}Erb, D. K., Steidel, C. C., Shapley, A. E.,
Pettini, M., Reddy, N. A., \& Adelberger, K. L. 2006b, \apj, 646, 107
\bibitem[Erb et al.(2006c)]{erb06c}Erb, D. K., Steidel, C. C., Shapley, A. E., Pettini, M., 
Reddy, N. A.,\& Adelberger, K, L. 2006c, \apj, 647, 128
\bibitem[Fan et al.(2001)]{fan01}Fan, X., et al., 2001, \aj, 121, 54
\bibitem[Finkelstein et al.(2009)]{fink09}Finkelstein, S. L., Papovich, C., Rudnick, G., Egami, E., Le Floc'h, E., 
Rieke, M. J., Rigby, J. R., \& Willmer, C. N. A., 2009, \apj, 700, 376
\bibitem[F\"orster Schreiber et al.(2009)]{fors09} F\"orster Schreiber, N. M. et al.
2009, \apj, 706, 1364
\bibitem[Hainline et al.(2009)]{hain09} Hainline, K. N., Shapley, A. E., 
Kornei, K. A., Pettini, M., Buckley-Geer, E., Allam, S. S., \& Tucker, D. L. 2009, \apj, 701, 52
\bibitem[Hill et al.(2008)]{hill08}Hill, J. M., Green, R. F., Slagle, J. H., Ashby, D. S., 
Brusa-Zappellini, G., Brynnel, J. G., Cushing, N. J., Little, J. \& Wagner, R. M., 2008, Proc. SPIE, 7012, 2
\bibitem[Kauffmann et al.(2003)]{kauf03}Kauffmann, G. et al. 2003, \mnras, 346, 1055
\bibitem[Kennicutt et al.(1998)]{kenn98}Kennicutt, R. C. 1998, \araa, 36, 189
\bibitem[Kewley et al.(2001)]{kewl01}Kewley, L. J., Dopita, M. A., Sutherland, R. S., Heisler, C. A.,
\& Trevena, J. 2001, \apj, 556, 121
\bibitem[Kim \& Koo(2001)]{kim01}Kim, K.-T.,\& Koo, B.-C. 2001, \apj, 549, 979
\bibitem[Kubo et al.(2009)]{kubo09}Kubo, J. M., Allam, S. S., Annis, J, Buckley-Geer, E. J., Diehl, H. T., 
Kubik, D., Lin, H., \& Tucker, D. 2009,\apj, 696, L61
\bibitem[Koester et al.(2010)]{koes10}Koester, B. P., Gladders, M. D., Hennawi, J. F., Sharon, K., Wuyts, E., Rigby, J. R., 
Bayliss, M. B., \& Dahle, H. 2010, arXiv:1003.0030
\bibitem[Lin et al.(2009)]{lin09}Lin, H., Buckley-Geer, E., Allam, S. S.; Tucker, Douglas L., Diehl, H. T., Kubik, D., 
Kubo, J. M., Annis, J., Frieman, J. A., Oguri, M., Inada, N. 2009, \apj, 699, 1424
\bibitem[Liu et al.(2008)]{liu08}Liu, X., Shapley, A. E., Coil, A. L., Brinchmann, J. E., \& Ma, C.-P. 2008,
\apj, 678, 758
\bibitem[Madau et al.(1998)]{madau98}Madau, P., Pozzetti, L., \& Dickinson, M. 1998, \apj, 98, 106
\bibitem[Mannucci et al.(2010)]{man10}Mannucci F., Cresci G., Maiolino R., Marconi A., \& Gnerucci, A., 2010,
arXiv:1005.0006
\bibitem[Osterbrock \& Ferland(2006)]{oste06}Osterbrock, D. E., \& Ferland, G. J. 2006, Astrophysiscs of 
Gaseous Nebulae and Active Galactic Nuclue, Second Edition (University Science Book) 
\bibitem[Mandel et al.(2008)]{mand08}Mandel, H. et al, 2008, Proc. SPIE, 4917, 124
\bibitem[Oesch et al.(2010)]{oesch10}Oesch, P. A., Bouwens, R. J., Carollo, C. M., Illingworth, G. D., Magee, D., 
Trenti, M., Stiavelli, M., Franx, M., \& Labbe, I., 2010, arXiv:1005.1661
\bibitem[Peng et al.(2002)]{peng02}Peng, C. Y., Ho, L. C.. Impey C. D., \& Rix, H., 2002, \aj, 124, 266
\bibitem[Peng et al.(2010)]{peng10}Peng, C. Y., Ho, L. C.. Impey C. D., \& Rix, H., 2010, \aj, 139, 2097
\bibitem[Peng et al.(2006)]{peng06}Peng, C. Y.. Impey, C. D., Rix, H., Kochanek, C. S., Keeton, C. R.,
Falco, E. E.; Leh\'ar, J., \& McLeod, B. A., 2006, \apj, 649, 616
\bibitem[Pettini \& Pagel(2004)]{pett04}Pettini, M. \& Pagel, B. E. J. 2004, \mnras, 348, L59
\bibitem[Pettini et al.(2001)]{pett01}	Pettini, M., Shapley, A. E., Steidel, C. C., 
Cuby, J., Dickinson, M., Moorwood, A. F. M., Adelberger, K. L., Giavalisco, M., 2001, \apj, 554, 981
\bibitem[Pettini et al.(2010)]{pett10}Pettini, M. et al. 2010, \mnras, 402, 2335
\bibitem[Press et al. (1992)]{press92}Press, W. H., Teukolsky, S. A., Vetterling, W. T., \& Flannery, B. P.
1992, Numerical recipes in C. The art of scientific computing
(Cambridge: University Press, â1992, 2nd ed.)
\bibitem[Quider et al.(2009)]{quid09}Quider, A. M., Pettini, M., Shapley, A. E.,\& Steidel, C. C. 
 2009, \mnras, 398, 1263
\bibitem[Quider et al.(2010)]{quid10}Quider, A. M., Shapley, A. E., Pettini, M., Steidel, C. C., 
\& Stark, D. P. 2010, \mnras, 402, 1467
\bibitem[Raimann et al.(2000)]{raim00}Raimann D, Storchi-Bergamann, T., Bica E., Melnick J., \& Schimitt, H., 2000, \mnras, 316, 559
\bibitem[Reddy et al.(2008)]{redd08}Reddy, N. A., Steidel, C. C., Pettini, M., Adelberger, K. L.,
Shapley, A. E., Erb, D. K., \& Dickinson, M. 2008, \apjs, 175, 48
\bibitem[Reddy \& Steidel(2009)]{redd09}Reddy, Naveen A. \& Steidel, Charle, C, 2009, \apj, 682, 778
\bibitem[Richard et al.(2008)]{richard08} Richard, J., Stark, 
D.~P., Ellis, R.~S., George, M.~R., Egami, E., Kneib, J.-P., \& Smith, G. P.\ 2008, \apj, 685, 705 
\bibitem[Sand et al.(2005)]{sand05} Sand, D.~J., Treu, T., 
Ellis, R.~S., \& Smith, G.~P.\ 2005, \apj, 627, 32 
\bibitem[Shapley et al.(2005)]{shap05}Shapley, A.~E., Steidel, C.~C., Erb, D.~K., Reddy, N.~A., Adelberger, K.~L., Pettini, M., Barmby, \& Huang, J., 2005, \apj,
626, 698
\bibitem[Smail et al.(2007)]{smai07}Smail, I., Swinbank, A. M., Richard, J., 
Ebeling, H., Kneib, J.-P., Edge, A. C., Stark, D.,
Ellis, R. S., Dye, S., Smith, G. P., \& Mullis, C. 2007, \apj, 654, 33
\bibitem[Storchi-Bergmann et al.(1994)]{stor94}Storchi-Bergamann, T., Calzetti D., \& Kinney A. L.,
1994, \apj, 429, 572 
\bibitem[Teplitz et al.(2000)]{teplitz00} Teplitz, H.~I., et al.\ 
2000, \apjl, 533, L65 
\bibitem[Yee et al.(1996)]{yee96} Yee, H.~K.~C., Ellingson, E., Bechtold, J., Carlberg, R.~G., 
\& Cuillandre, J.-C.\ 1996, \aj, 111, 1783 
\bibitem[York et al.(2000)]{york00}York, D. G., et al. 2000, \aj, 120, 1579
\bibitem[Zaritsky et al.(1994)]{zari94}Zaritsky, D., Kennicutt, R. C., \& Huchra, J. P., 1994, \apj, 420, 87
\end{thebibliography}
\end{document}